\documentclass[aps,superscriptaddress,eqsecnum,nofootinbib,showpacs,preprintnumbers]{revtex4-2}
\usepackage[utf8]{inputenc}
\usepackage{amsmath}
\usepackage{amssymb}
\usepackage{amsfonts,amsthm,bm}
\usepackage{color,xcolor}
\usepackage{comment}
\usepackage{soul}
\usepackage[commandnameprefix=always]{changes}
\usepackage{graphicx,epsfig}
\usepackage{float}
\usepackage{subcaption}
\usepackage{tikz}
\usepackage{stmaryrd}

\newcommand{\be}{\begin{eqnarray}}
	\newcommand{\ee}{\end{eqnarray}}
\newcommand{\bea}{\begin{eqnarray}}
	
	\newcommand{\eea}{\end{eqnarray}}

\usepackage{ulem}

\definecolor{azure(colorwheel)}{rgb}{0.0, 0.5, 1.0}
\definecolor{DarkViolet}{RGB}{148,0,211}
\definecolor{myDarkBlue}{rgb}{0,0.1,0.7}
\definecolor{DarkBlue}{RGB}{0,0,153}
\definecolor{amber}{rgb}{1.0, 0.49, 0.0}
\definecolor{amaranth}{rgb}{0.9, 0.17, 0.31}
\definecolor{nicered}{rgb}{0.7,0.1,0.1}
\definecolor{brown}{rgb}{0.5,0.1,0.1}
\definecolor{nicegreen}{rgb}{0.0,0.3,0.0}
\definecolor{tealgreen}{rgb}{0.0, 0.51, 0.5}

\definecolor{tclr}{RGB}{148,0,211}

\usepackage{hyperref}
\hypersetup{colorlinks,bookmarksopen,
	bookmarksnumbered,
	citecolor={nicered},
	linkcolor={myDarkBlue},
	urlcolor={tealgreen},
	pdfstartview=FitH}

\newcommand{\beq}{\begin{equation}}
\newcommand{\eeq}{\end{equation}}
\newcommand{\bseq}{\begin{subequations}}
	\newcommand{\eseq}{\end{subequations}}

\usepackage{times}

\graphicspath{{Figs/}}

\usepackage{orcidlink}
\def\idsari{\orcidlink{0000-0003-1500-0874}}
\def\idziyu{\orcidlink{0000-0003-3039-9975}}
\def\idkara{\orcidlink{0000-0002-5479-6513}}

\usepackage{bigints}

\begin{document}

\title{Thermodynamically massless Simpson-Visser black holes}

\author{Thanasis Karakasis\idkara}
	\email{thanasiskarakasis@mail.ntua.gr (Corresponding author)}
	\affiliation{Physics Department, School of Applied Mathematical and Physical Sciences,
	National Technical University of Athens, 15780 Zografou Campus,
	Athens, Greece.}

\author{Emmanuel N. Saridakis\idsari}

\affiliation{Institute for Astronomy, Astrophysics, Space Applications and Remote Sensing,
National Observatory of Athens, 15236 Penteli, Greece}    
\affiliation{Departamento de Matem\'{a}ticas, Universidad Cat\'{o}lica del 
Norte, Avda. Angamos 0610, Casilla 1280 Antofagasta, Chile}
\affiliation{CAS Key Laboratory for Researches in Galaxies and Cosmology, 
School 
of Astronomy and Space Science, University of Science and Technology of China, 
Hefei, Anhui 230026, China}

\author{Zi-Yu Tang\idziyu}
\email{tangziyu@ibs.re.kr (Corresponding author)}

\affiliation{Cosmology, Gravity and Astroparticle Physics Group, Center for Theoretical Physics of the Universe, Institute for Basic Science, Daejeon 34126, Korea}

\begin{abstract}

In this work, we scrutinize the thermodynamic properties of the Simpson-Visser (SV) spacetime. Working within Einstein gravity coupled to nonlinear electrodynamics (NLED) and a scalar field with negative kinetic energy, we rederive the solution in a formulation where the integration constants do not explicitly appear in the action, allowing them to vary consistently in the thermodynamic analysis. Using the Euclidean method, we show that the regular spacetime structure modifies the boundary contributions to the conserved charge associated with time translations, allowing the NLED sector to cancel the mass term and yielding a black hole with vanishing thermodynamic mass. Nevertheless, the spacetime admits a conserved magnetic charge and describes a regular black hole with a single horizon, finite temperature, and entropy, while the first law of thermodynamics holds. We further compare this solution with the corresponding scalar-free singular black hole obtained when the regular parameter vanishes. Placing the two configurations in the same heat bath with identical temperature and magnetic chemical potential, we find that the SV regular black hole always has a larger free energy, indicating that the scalar-free singular configuration is thermodynamically preferred. 

\end{abstract}

\maketitle

\section{Introduction}

Black holes constitute one of the most remarkable predictions of General 
Relativity (GR). Despite their success in describing gravitational phenomena 
over a wide range of scales, classical black-hole solutions typically possess 
spacetime singularities, where curvature invariants diverge and the classical 
description of gravity breaks down. The presence of such singularities reveals 
the limits of the theory and indicates that a more fundamental framework, 
likely incorporating quantum gravitational effects, is required in order to 
describe the extreme regions of spacetime. 

Regular black holes are black-hole spacetimes that are devoid of essential 
singularities.
Essential singularities are points where curvature invariants, such as the Ricci 
or Kretschmann scalar, diverge and where null and timelike geodesics 
terminate.  These 
singularities differ from coordinate singularities, since they cannot be 
removed by an appropriate coordinate transformation. As a result, the 
effective theory from which such a spacetime arises as a solution loses 
predictability there and must be replaced by a more appropriate quantum 
description. This is the case of GR and the well-known Schwarzschild solution, 
where an essential singularity exists at the origin of the coordinate system 
($r=0$). By constructing black-hole solutions with smooth cores, one may gain 
insight into how quantum gravitational effects could modify the structure of 
spacetime in extreme regimes. In this sense, regular black holes serve as 
valuable ``toy models'' that can guide the search for a more complete theory of 
gravity, such as loop quantum gravity or string theory
\cite{Calza:2024fzo, Calza:2024xdh, Calza:2024qxn,  Calza:2025mwn, Calza:2025yfm, Vagnozzi:2022moj,bardeen1968,Borde:1996df,Burinskii:2002pz,Dymnikova:2004zc,
Hayward:2005gi,
Bronnikov:2006fu,Berej:2006cc,
Lemos:2011dq,Flachi:2012nv,Li:2013jra,Neves:2014aba,
Bambi:2013ufa, 
Abdujabbarov:2016hnw, Toshmatov:2017zpr,Nashed:2018cth, 
Jusufi:2018jof,Carballo-Rubio:2018pmi, 
Simpson:2019mud,Kumar:2019pjp,Nashed:2020kdb, Babichev:2020qpr,
Ovgun:2019wej, Nashed:2021pah, Jusufi:2022rbt,   Nashed:2022yfc,Bueno:2024dgm, 
Cisterna:2025vxk,Nashed:2025bxv, Bakopoulos:2023fmv} (for a review see  
\cite{Lan:2023cvz}).  
 
Since a full quantum theory of gravity is still absent, these solutions are 
usually studied at the classical level in order to understand their physical 
properties and their possible relation to quantum gravity through 
semiclassical considerations, such as black-hole thermodynamics. The first 
regular black-hole spacetime was proposed by Bardeen \cite{bardeen1968}, and it was 
later realized that this solution can arise from Einstein gravity coupled to a 
particular nonlinear electrodynamics (NLED) theory \cite{Ayon-Beato:2000mjt}. 
Over the years, a general procedure has been developed for constructing regular 
black-hole solutions supported by NLED 
\cite{Pellicer:1969cf,Bronnikov:2000vy,Fan:2016hvf,Cisterna:2020rkc, 
Barrientos:2022bzm, Murk:2024nod,Barrientos:2024umq,Barrientos:2025rjn}.  In 
this 
approach one first specifies the desired regular spacetime geometry and then 
solves the resulting Einstein equations in order to reconstruct the NLED 
theory capable of supporting it. This scenario, however, presents several 
drawbacks. In particular, the resulting NLED Lagrangian may take an exotic 
form, or the conserved black-hole parameters may appear explicitly in the 
Lagrangian. In that case the parameters of the solution, such as the mass $M$ 
and the electric or magnetic charge $Q$, are promoted to theory constants and 
are therefore not allowed to vary at the classical level, leading to 
deviations from the standard first law of black-hole thermodynamics. Finally, 
regular 
black-hole solutions have also been obtained in various modified 
gravity theories, where higher-curvature corrections, scalar degrees of 
freedom, or alternative gravitational dynamics can effectively regularize 
the central region of the spacetime 
\cite{Cognola:2011nj, Myung:2013doa, Bambi:2015kza, 
Ali:2018boy,Carballo-Rubio:2021wjq,
 Junior:2023ixh, Bakopoulos:2023sdm, Babichev:2023mgk, 
Nozari:2023enj, Luciano:2023bai, 
Johannsen:2013szh,  Boos:2024sgm,  Konoplya:2023ppx, 
 Nozari:2024vxp,Estrada:2024uuu,      Bueno:2025gjg, 
Charmousis:2025xug,
 Santos:2025fdp,
  Nozari:2025rkc, Anand:2025cer, Nozari:2026wjo, Sucu:2026nkw}.

In this work we focus on the Simpson-Visser (SV) spacetime 
\cite{Simpson:2018tsi,Simpson:2019cer,Lobo:2020ffi, Tsukamoto:2020bjm}, which provides a simple 
deformation of the Schwarzschild geometry through the introduction of a new 
length scale $\alpha$. This modification effectively replaces the radial 
coordinate $r$ with $\sqrt{r^2+\alpha^2}$, transforming the singular point at 
$r=0$ into a regular two-sphere of radius $\alpha$. As a result, the spacetime 
becomes smooth at the origin and interpolates between different geometric 
configurations, including a regular black hole with a spacelike throat and 
wormhole geometries with timelike or null throats. The parameter $\alpha$ thus 
acts as a core scale that regularizes the central region of the spacetime.

Following its introduction, several extensions of the SV geometry have been 
proposed, incorporating electric charge and rotation 
\cite{Franzin:2021vnj,Mazza:2021rgq,Shaikh:2021yux,Guo:2021wid}. A closely 
related regularization mechanism had already been discussed by Bronnikov and 
Fabris \cite{Bronnikov:2005gm}, where a phantom scalar field with a 
self-interacting potential naturally leads to the replacement 
$r\to\sqrt{r^2+\alpha^2}$ in the angular sector of the metric. In this 
framework the parameter $\alpha$ is associated with the scalar charge and 
controls the asymptotic behavior of the field, giving rise to a secondary 
scalar hair \cite{Karakasis:2023hni}. The explicit theory supporting the SV 
geometry was later constructed in \cite{Bronnikov:2021uta}, where Einstein 
gravity is coupled to a particular nonlinear electrodynamics sector and a 
self-interacting phantom scalar field. The scalar sector plays a 
crucial role in the regularization mechanism, replacing the central singularity 
with a smooth core supported by the matter fields.

In this work we investigate the thermodynamic properties of the uncharged 
Simpson-Visser (SV) spacetime when it arises as a solution of the theory 
constructed in \cite{Bronnikov:2021uta}. We first re-derive the solution in a 
form where the spacetime parameters do not appear as coupling constants of the 
underlying theory. In this formulation the SV geometry is described by a single 
free integration constant that can be consistently varied in the thermodynamic 
analysis. 

Using Euclidean methods and the formalism of Gibbons and Hawking 
\cite{Gibbons:1976ue}, we compute the Euclidean action and construct the first 
law of thermodynamics for this system. We show that the requirement of a 
well-defined variational principle leads to a remarkable result: the 
contribution of the nonlinear electrodynamics sector exactly cancels the pure 
gravitational contribution to the conserved mass. As a consequence, the SV 
solution describes thermodynamically massless black holes, even though the 
spacetime still contains a non-vanishing geometric mass parameter. This implies 
that the thermodynamic mass obtained from the Euclidean action differs from 
other notions of mass, such as the Komar mass, and highlights how nonlinear 
matter couplings can modify the standard first law of black-hole thermodynamics.

The structure of the paper is as follows. In Sec. \ref{Theory} we review the 
theory 
that 
supports the SV spacetime and present the corresponding solution. In Sec 
\ref{Reconstruction} 
we reconstruct the theory in a form where the spacetime parameters appear as 
integration constants rather than coupling constants. In Sec. \ref{thermo} we 
analyze 
the 
thermodynamics of the SV geometry using the Euclidean action formalism and 
derive the modified first law. In Sec. \ref{Scalarfree} we examine the 
scalar-free 
configuration and compare its thermodynamic behavior with that of the SV 
solution. Finally, in Sec. \ref{Conclusions} we summarize our results and 
discuss their 
implications.

\section{The theory and the Simpson-Visser  solution}
\label{Theory}

In this section we briefly review the field-theory framework that supports the 
Simpson-Visser (SV) spacetime. In particular, we present the gravitational 
theory consisting of Einstein gravity coupled to a phantom scalar field and a 
non-linear electromagnetic sector, as constructed in \cite{Bronnikov:2021uta}. 
Within this theory the SV geometry arises as an exact solution of the field 
equations. Introducing the action and the corresponding field equations will 
allow us to identify the role played by the various fields and parameters of 
the solution, which will be important for the thermodynamic analysis that 
follows.

The theory supporting the SV spacetime \cite{Bronnikov:2021uta} can be written 
in $\left(-+++\right)$ signature as
\begin{equation}
    S =\frac{1}{2}\int d^4x \sqrt{-g}\Big\{R + 
2\nabla_{\mu}\phi\nabla^{\mu}\phi - V(\phi) - \mathcal{L}(
\mathcal{F})\Big\}~, \label{generalaction}
\end{equation}
where $R$ is the Ricci scalar, $\mathcal{F} = F_{\mu\nu}F^{\mu\nu}$, 
$F_{\mu\nu} = \partial_{\mu}A_{\nu} - \partial_{\nu}A_{\mu}$ and $\phi$ is a 
phantom scalar field with negative kinetic energy. The field equations with 
respect to the dynamical fields ($g_{\mu\nu}, A_{\mu}, \phi$) read
\begin{eqnarray}
    &&G_{\mu\nu}  = T_{\mu\nu}^{EM} + T_{\mu\nu}^{\phi}~, 
\label{fieldequations}\\
    &&T_{\mu\nu}^{EM} = 
2\mathcal{L}_{\mathcal{F}}F_{\mu\sigma}F_{\nu}^{~\sigma} - 
\frac{1}{2}g_{\mu\nu}\mathcal{L}(\mathcal{F})~,\\
    &&T_{\mu\nu}^{\phi} = -2\nabla_{\mu}\phi\nabla_{\nu}\phi + 
g_{\mu\nu}\left(\nabla^{\sigma}\phi\nabla_{\sigma}\phi - 
\frac{1}{2}V(\phi)\right)~,\\
    &&\nabla_\sigma \nabla^\sigma \phi = -\frac{1}{4}V'(\phi)~,\\
    &&\nabla_{\mu}(\mathcal{L}_{\mathcal{F}} 
F^{\mu\nu})=0~,\label{fieldequations1}
\end{eqnarray}
where we have set Einstein's constant $\kappa\equiv 8\pi G=1$. The form of 
scalar potential and non-linear electromagnetic Lagrangian that allow the 
existence of the SV spacetime are given by \cite{Bronnikov:2021uta}
\begin{eqnarray}
    &&V(\phi) = \frac{2m}{5\alpha^3} \cos^5\phi ~,\\
    &&\mathcal{L}(\mathcal{F}) = \frac{3\ 2^{3/4} m \alpha ^2 \mathcal{F}^{5/4} 
}{5 Q_m^{5/2}}~,
\end{eqnarray}
and the full solution to the field equations reads
\begin{eqnarray}
    &&ds^2 = -F(r)dt^2 + \frac{dr^2}{F(r)} + w^2(r) d\Omega^2~, 
\label{metric_ansatz}\\
    &&F(r) = 1 - \frac{2m}{\sqrt{r^2+\alpha^2}} ~, \label{sol1}\\
    &&w(r) = \sqrt{r^2+\alpha^2}~,\\
    &&\phi(r) = \arctan \frac{r}{\alpha}~,\\
    &&A_{\mu} = A_3(\theta) d \varphi = Q_m \cos \theta d\varphi~. \label{sol2}
\end{eqnarray}
By incorporating additional terms in the scalar potential and the NLED 
Lagrangian, the electromagnetically charged SV spacetime may also be supported 
\cite{Bronnikov:2021uta}, but we will not deal with this scenario here. 
Here, $m$ represents the black-hole mass, $Q_m$ the magnetic charge and 
$\alpha$ is a deformation parameter that renders all quantities finite at the 
origin $r=0$. Notice that $r=0$ now reduces to a 2-sphere of finite radius 
$\alpha$ instead of a point as in the Schwarzschild case where $\alpha=0$. 
Hence, the ``origin" of this black hole can be spacelike, timelike or null. 
This is the reason behind the finiteness of all curvature invariants at $r=0$. 
In fact, the  $r=0$ hypersurface corresponds to a wormhole throat, which in 
general is located at the minimum of the $g_{\theta\theta}$ in this coordinate 
system \cite{Morris:1988cz}. In this spacetime we have $g_{\theta\theta}'(r=0) 
=0 ~,g_{\theta\theta}''(r=0)>0$, hence at $r=0$ we have the location of the 
throat, with its radius given by $r_{\text{throat}} 
=\sqrt{g_{\theta\theta}(r=0)} = |\alpha|$. Now, as one can see, the constants 
$m,Q_m$ and $\alpha$ are included in the Lagrangian of the theory, which is a 
common scenario in the reconstruction method, where the geometry is imposed, 
and the theory that can support the geometry can be reconstructed. As a result, 
the integration constants of the black hole $m,Q_m, \alpha$ have been promoted 
to constants of the theory and cannot be varied at the classical level. In the 
next subsection we will show that it is possible to derive a similar or same 
line element to the original SV spacetime, where the integration constants do 
not appear in the Lagrangian of the theory, we will therefore re-write the 
solution and theory of \cite{Bronnikov:2021uta} in a field-theory consistent 
form.

\section{Reconstruction with variable integration constants}
\label{Reconstruction}

In this section we reconstruct the SV spacetime in a form that allows for a 
consistent thermodynamic interpretation. As discussed in the previous section, 
in the original reconstruction approach the black-hole parameters may appear 
explicitly in the Lagrangian of the theory. In such a case these parameters 
become coupling constants of the theory and therefore cannot vary at the 
classical level. To study the thermodynamics of the spacetime
it is desirable to formulate the theory in such a way that the parameters of 
the solution arise as integration constants of the field equations. 

To achieve this, we will re-derive the SV solution starting from the action 
(\ref{generalaction}) while keeping the couplings of the theory independent of 
the integration constants of the solution. We will show that the SV spacetime 
can indeed be written in such a way that it is described by a free parameter 
that is allowed to vary. Once this is established, we will then use first 
principles to derive the first law of thermodynamics for this spacetime.

To re-derive the solution and examine whether it is possible to express the 
solution in such a way that the parameters do not explicitly appear in the 
Lagrangian, we begin with the theory (\ref{generalaction}) and the following 
scalar potential and NLED Lagrangian
\begin{eqnarray}
    V(\phi) &=& h_1 \cos ^5 \phi~, \label{potential}\\
    \mathcal{\mathcal{L}}\left(\mathcal{F}\right) &=& 
f_1\mathcal{F}^{5/4}~,\label{LF}
\end{eqnarray}
where $h_1~, f_1$ are constants of the theory with appropriate units. In 
particular, $h_1$ has units of $[L]^{-2}$, while $f_1$ has units of 
$[L]^{1/2}$. For the line element we consider the following spacetime metric
\begin{equation}
     ds^2 = -N(r)^2 F(r) dt^2 + \frac{dr^2}{F(r)} + w(r)^2 d\Omega^2~, 
\label{metric}
\end{equation}
with $N(r)$ being an auxiliary function. This function can be reabsorbed by a coordinate transformation and is therefore not fixed by the field equations. Nevertheless, in the next subsection, where we discuss the thermodynamic aspects, it facilitates a simpler computation of the effective Lagrangian. For completeness, we thus introduce it here. For the electromagnetic sector of the 
action we consider a radial magnetic field (see Appendix \ref{AppendixA}) with 
vector potential
\begin{equation}
    A_\mu =\left(0,0,0, A_3(\theta)\right)~, \label{A}
\end{equation}
which leads to the only non-vanishing components 
$F_{\theta\varphi}=-F_{\varphi\theta}=A_3'(\theta)$. The field equations 
(\ref{fieldequations})-(\ref{fieldequations1}) lead to 
\begin{eqnarray}
    &&\frac{2 }{w^2}\left(w  \left(F'  w' +2 F  w'' \right)+F  
w'{}^2-1\right)-2 F  \phi' {}^2+V(\phi)+\mathcal{L}\left(\mathcal{F}\right)=0~, 
\label{eq0}\\
    &&\frac{w \left(\mathcal{L}\left(\mathcal{F}\right)+V(\phi)\right)+2 F'  w' 
-\frac{2}{w }}{F}+\frac{4 N' w'}{N}+\frac{2 w' {}^2}{w }+2 w  \phi ' 
{}^2=0~,\label{eq1}\\
    &&N w \mathcal{L}\left(\mathcal{F}\right)+3 w F'N' +N \left(w F''+2 F'  
w'+2 F w''-2 F w \phi' {}^2+w V(\phi)\right)+2 F \left(w  N'\right)' =\frac{4 
\csc^2{\theta}  A_3 ' {}^2 \mathcal{L}'(\mathcal{F})N}{w^3}~, \label{eq2}\\
    &&4 F'  \phi ' +4 F \left(\frac{N' \phi '}{N }+\frac{2 w' \phi' }{w }+\phi 
'' \right)+V'(\phi)=0~, \label{eqKG}\\
    &&\cot{\theta}   A_3'(\theta) -A_3''(\theta) =0~, \label{eqF}
\end{eqnarray}

where primes without specific arguments denote derivatives with respect to $r$. Due to the Bianchi identity, the Klein-Gordon equation (\ref{eqKG}) can be derived from the Einstein equations (\ref{eq0})-(\ref{eq2}) and is therefore redundant.
The equation (\ref{eqF}) gives 
\begin{equation}
    A_3(\theta)= Q_m \cos{\theta}~, \label{A3}
\end{equation}
where $Q_m$ is the magnetic charge. Note that $\mathcal{F}=2 \csc ^2(\theta ) 
A_3'(\theta )^2/w (r)^4$, after substituting (\ref{A3}) it becomes only 
$r$-dependent, $\mathcal{F}=2Q_m^2/w(r)^4$. Conversely, if we require 
$\mathcal{F}$ to be only $r$-dependent then it leads to the same solution 
(\ref{A3}) and the electromagnetic field equation (\ref{fieldequations1}) is 
satisfied automatically. A Coulomb-like radial magnetic field $B^r\sim Q_m/r^2$ 
also implies the same form $A_3(\theta)=Q_m \cos{\theta}$.

Equations (\ref{eq0}) and (\ref{eq1}) lead to the simple 
relation
\begin{equation}
    \frac{N' w'}{N}+w \phi' {}^2=w'' . \label{eq01}
\end{equation}
With $N(r)=const.$ and $w(r) = \sqrt{r^2+\alpha^2}$, the scalar-field profile 
becomes 
\begin{equation}
    \phi(r) = \pm\arctan \left(\frac{r}{\alpha}\right)+\phi_0~,
\end{equation}
where we set $\phi_0=0$ in order to obtain an asymptotically flat spacetime 
with $V(\phi)\to 0$ at spatial infinity. Without loss of generality we choose 
the ``$+$" branch. In principle, the field equations can determine all the functions, including $w$ and $\phi$. In this work, however, we focus on the SV case, for which this choice of $w$ leads to the constraints (\ref{relation}) required for the compatibility of the field equations. By contrast, choosing $w=r$ yields a constant scalar field and a singular black hole, as discussed in Sec. \ref{Scalarfree}.

Thereafter the metric function can be obtained from the Klein-Gordon equation 
(\ref{eqKG}),
\begin{equation}
    F(r)=c_1-\frac{5\alpha^3 h_1}{4\sqrt{r^2+\alpha^2}}~,
\end{equation}
where $c_1$ is an integration constant. In order to satisfy the remaining 
equations the following relations must hold for positive $\alpha$
\begin{equation}
    c_1=1~, \quad \frac{h_1}{f_1}=\frac{2  \left(2Q_m^2\right)^{5/4}}{3 
\alpha^5}~, \label{relation}
\end{equation}
where $h_1$ and $f_1$ have the same sign. The second relation implies that the 
ratio $Q_m/\alpha^2$ is fixed in terms of the coupling constants $f_1,~h_1$, 
namely
\begin{equation}
    \left(\frac{Q_m}{\alpha^2}\right)^2=\frac{1}{2}\left(\frac{3 
h_1}{2f_1}\right)^{4/5}~.\label{Qm2}
\end{equation}

For negative $\alpha$ the relations obtained from different equations become 
contradictory, unless $h_1=0$ and $f_1 \times Q_m=0$. Therefore in what follows 
we consider only positive $\alpha$.
The asymptotic expression of the metric function $F$ at large distances is 
\begin{equation}
    F(r\to\infty) \sim 1 -\frac{5 \alpha^3 h_1}{4 r} + \mathcal{O}(r^{-3})~,
\end{equation}
from which we see that the core scale $\alpha$ also determines the effective 
Schwarzschild mass. In the weak-field limit we may identify the Newtonian 
gravitational potential through $g_{tt} \sim -1 + 2\Phi_N$, with $\Phi_N = 
5\alpha^3 h_1 /(8 r)$. The mass of the configuration is therefore $M = 
5\alpha^3 h_1 /8$. For completeness, we also compute it using the Komar method, as follows.

The spacetime admits an asymptotically timelike Killing vector field $K^{\mu} = 
(1,0,0,0)$ with dual $K_{\mu} = -F(r) dt$. From this we construct the two-form 
$X=dK=-F'(r)dr\wedge dt$, whose Hodge dual is $\star X=-F'(r) w^2(r) \sin \theta d\theta\wedge 
d\varphi$. The Komar mass then reads
\begin{equation}
    M_{\text{Komar}} = -\frac{1}{4\pi} \lim_{r\to \infty} \int_{S} \star X = 
-\frac{1}{4\pi}\lim_{r\to \infty} \left(\frac{5 \pi  \alpha ^3 h_1 r}{2 
\sqrt{\alpha ^2+r^2}}\right) = \frac{1}{8\pi} \left(5 \left(\pi  \alpha ^3 
h_1\right)\right) = M~,
\end{equation}
and therefore the Komar mass is determined by the integration constant $\alpha$ 
and the theory constant $h_1$. In order to obtain a positive mass we require 
$h_1>0$. The existence of the timelike Killing vector thus leads to a conserved 
geometric mass. However, as we will show in the following subsection, 
semiclassical arguments imply that this mass does not coincide with the 
thermodynamic mass derived from the Euclidean partition function of the black 
hole.

The full solution of the theory (\ref{generalaction}) reads
\begin{eqnarray}
    &&F(r) = 1-\frac{5 \alpha ^3 h_1}{4 \sqrt{\alpha ^2+r^2}}~, \label{F} \\ 
    &&w(r) = \sqrt{r^2+ \alpha^2}~, \label{w}\\
    &&A_3(\theta) = \pm \frac{\alpha^2}{\sqrt{2}}\left(\frac{3 
h_1}{2f_1}\right)^{2/5} \cos \theta~,\\
    &&\phi(r) = \arctan \left(\frac{r}{\alpha}\right)~. \label{phi}
\end{eqnarray}
From these relations it becomes clear that the solution possesses a genuine 
free parameter, namely the core scale $\alpha$, which is allowed to vary. The 
theory parameters $f_1$ and $h_1$ are independent but must have the same sign. If the function $w(r)$ were not fixed, the solution could instead describe a singular black hole, as discussed above. In that case, the mass and magnetic charge are independent, and the parameter space could be enlarged. In contrast, for the SV spacetime, the only free parameter is the core scale $\alpha$, related to the mass and magnetic charge.

Furthermore, when $h_1=0$ the scalar potential $V(\phi)$ vanishes, as does the 
magnetic potential $A_3$, which results in the disappearance of the NLED sector 
of the action. In this limit the theory reduces to the one supporting the Ellis 
drain-hole solution \cite{Ellis:1973yv, Karakasis:2021tqx,Bakopoulos:2023tso}. 
The inclusion of the scalar potential therefore plays a crucial role in 
allowing the existence of a black-hole solution.

 \begin{figure}[ht]
     \centering
     \includegraphics[width=0.55\linewidth]{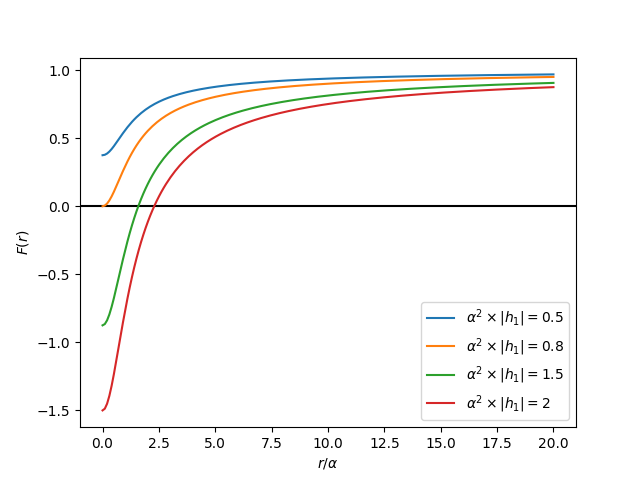}
\caption{{\it{Metric function $F(r)$ as a function of the dimensionless radius $r/\alpha$ for different values of $\alpha^2|h_1|$. The orange line represents the extremal case, where the horizon coincides with the wormhole throat.}}}
     \label{fig:placeholder}
 \end{figure}

For $\alpha^2>4/(5|h_1|)$ the spacetime contains a black-hole horizon located at
\begin{equation}
     r_h = \frac{\alpha}{4} \sqrt{25 \alpha ^4 h_1^2-16}~,
\end{equation}
and in FIG. \ref{fig:placeholder} we plot the metric function $F(r)$. For $\alpha^2 < 4/(5h_1)$, no horizon is present. Nevertheless, the
spacetime remains regular, as the SV geometry is free of curvature singularities at $r=0$. In order to obtain a well-defined 
black-hole geometry the horizon radius must exceed the location of the wormhole throat, which is at $r=0$, and is realized when $\alpha^2>4/(5|h_1|)$~. As illustrated in 
FIG.~\ref{fig:placeholder}, the spacetime can therefore admit configurations 
with a horizon larger than the location of the wormhole throat. The extremal configuration corresponds to 
the horizon coinciding with the wormhole throat, $r_h=0$, which occurs for $\alpha^2=4/(5|h_1|)$, i.e., $\alpha^2|h_1| = 4/5 = 0.8$. This case is also included in FIG.~\ref{fig:placeholder}. In 
this work we focus on the regime $\alpha^2 \ge 4/(5|h_1|)$ where the 
horizon is at least equal to the throat.  Even in the extremal case the 
spacetime preserves its black-hole character, and this configuration may be 
referred to as a ``black throat'' \cite{Bronnikov:2021uta}.

\section{Thermodynamics of the  Simpson-Visser  spacetime}  \label{thermo}

Having obtained a formulation of the SV spacetime where the relevant parameter 
appears as a genuine integration constant, we may now proceed to investigate 
its 
thermodynamic properties. In the semiclassical description of black holes, the 
thermodynamics of the system can be derived from the gravitational partition 
function using the Euclidean path–integral formalism developed by Gibbons and 
Hawking. In this framework the classical Euclidean action evaluated on-shell 
encodes the thermodynamic potentials of the spacetime.

To study the thermodynamics of the theory and the corresponding black-hole 
solution we therefore work in the Euclidean formulation. We consider a static 
Euclidean spacetime with metric
\begin{equation}
    ds^2 = N(r)^2 F(r) d\tau^2 + \frac{dr^2}{F(r)} + w(r)^2 d\Omega^2~, 
\label{adm}
\end{equation}
which is obtained through the Wick rotation $t \to -i \tau$. Under this 
transformation the Lorentzian action $S$ is related to the Euclidean action 
$\mathcal{I}_E$ through $i S \to -\mathcal{I}_E$~\cite{Bakopoulos:2024hah}. In the semiclassical 
approximation the thermodynamic properties of the system are encoded in the 
Euclidean partition function,
\begin{equation}
    \mathcal{Z} = \int d[g_{\mu\nu}^E,\psi]e^{-\mathcal{I}_E}~,
\end{equation}
where the subscript $E$ denotes Euclidean quantities and $\psi$ collectively 
represents the matter fields.
The spacetime element has a Euclidean signature $(+,+,+,+)$. If we are 
considering a black hole spacetime, then to avoid the conical singularity at 
the horizon of the black hole $r_h$, the time coordinate $\tau$ must be 
periodic, with a period given by \cite{Bakopoulos:2024hah,Karakasis:2023hni}
\begin{equation}
    \beta = \frac{4\pi}{N\left(r_h\right)F'\left(r_h\right)}~,\label{period}
\end{equation}
and now $0 \leq \tau \leq \beta,~ r_h\le r<\infty$, while the spherical angles 
assume their usual ranges. This periodicity will be associated with the inverse 
temperature of the black hole, and as one can see, the temperature of the black 
hole solution will be determined entirely from the spacetime metric. The free 
energy of a statistical system, $\mathcal{G}$, will be related to the partition 
function via \cite{Gibbons:1976ue}
\begin{equation}
    \mathcal{G} = -\ln \mathcal{Z}/\beta~. \label{euclid}
\end{equation}
By imposing the saddle point approximation, we may consider that for the black 
hole spacetime that will concern us in this work, it is sufficient to consider 
the contribution of the action of the theory evaluated on the classical 
solution when the field equations hold: $\delta \mathcal{I}_E = 0$~. 
Consequently, we need to calculate the on-shell action of our theory, ensuring 
it attains a true extremum within the class of fields considered. From this, we 
can relate the Euclidean action to the free energy of the system via
\begin{equation}
    \mathcal{I}_E = \mathcal{G}\beta~. \label{action_free_energy}
\end{equation}

The action (\ref{generalaction}) can be written in the following form
\begin{equation}
    \mathcal{I}_E =  2\pi \beta \int _{0}^{\pi} d\theta \int_{r_h}^{\infty} dr 
N\mathcal{H} + \mathcal{B}~. \label{eeaction}
\end{equation}
In this expression, 
 $\mathcal{H}$ is a quantity that acts as a constraint and vanishes when the 
field equations hold (on-shell), while $N$ acts as a Lagrange multiplier, and 
we have performed the integrations over the periodic coordinates $\tau$ and 
$\varphi$.
No momentum terms are included due to the fact that the scalar field and the 
spacetime are time-independent, and we only consider magnetic fields. 
$\mathcal{B}$ denotes collectively the boundary terms required to ensure a 
well-defined variational problem \cite{Bakopoulos:2024hah}. 
To compute the before-given action, we substitute the Euclidean line element 
(\ref{adm}), the gauge field (\ref{A}) in (\ref{eeaction}) and obtain that, 
\begin{equation}
\begin{split}
    \mathcal{H} =& \frac{1}{2} \sin (\theta ) \bigg(2 w\left(F' w'+2 F 
w''\right)+w^2 \left(h_1 \cos ^5\phi-2 F 
   \left(\phi '\right)^2\right)+2  \left(F \left(w'\right)^2-1\right)\bigg)
    +\frac{2 \sqrt[4]{2} f_1 \csc ^{\frac{3}{2}}(\theta ) \left(A_3'(\theta 
)^2\right)^{5/4}}{w^3}~,
   \end{split}\label{Hequation}
\end{equation}
which will vanish on shell and is equivalent to the field equation (\ref{eq0}).

The first terms of expression (\ref{Hequation}), denote the contributions 
from the gravity and the scalar field part of the action, while the last term is the contribution of the magnetic field and primes denote derivative 
with respect to the argument. All functions in this expression are 
$r$-dependent, with the only exception being $A_3$, which is $\theta$-dependent 
in order to generate a radial magnetic field, compatible with the spherical 
symmetry of our system. We would like to compute the Euclidean action 
(\ref{eeaction}), in order to compute the free energy of the black hole, 
(\ref{action_free_energy}), when the field equations hold. Variation of 
(\ref{eeaction}) with respect to the function $N$, yields that $\mathcal{H}=0$, 
and hence the Euclidean action (\ref{eeaction}) is determined completely by the 
boundary term $\mathcal{B}$ when the field equations hold: $\mathcal{I}_E = 
\mathcal{B}$. The solution reported in equations (\ref{F})-(\ref{phi}) satisfies 
the Euler-Lagrange equations emanating when varying (\ref{eeaction}) with 
respect to $F,w,\phi, A_3$ for $N=1$.

In order to derive the field equations we have canceled several boundary terms. 
The boundary term $\mathcal{B}$ has to have a particular form in order to 
compensate for these boundary terms so that we have a well-defined variational 
procedure $\delta \mathcal{I}_E=0$. Consequently, we will calculate the 
variation of the boundary term $\delta \mathcal{B}$ and through this we will 
obtain $\mathcal{B}$ and, hence, $\mathcal{I}_E$ \cite{Alkac:2024hvu, 
Bakopoulos:2024hah, Bakopoulos:2024ogt, Bakopoulos:2024zke, Bakopoulos:2025eps, 
Martinez:2004nb}. In the variation $\delta \mathcal{B}$, we find that the terms 
proportional to $\delta w$, $\delta F$, $\delta \phi$ and $\delta A_3$ are 
precisely multiplied by the corresponding field equations 
(\ref{eq2}),(\ref{eq01}), (\ref{eqKG}) and (\ref{eqF}). Therefore, these 
contributions vanish on shell, and we finally obtain
\begin{equation}
    \delta \mathcal{I}_E=4\pi \beta \left[ww' \delta F-wF'\delta w+2wF \delta 
w'-2F w^2 \phi' \delta\phi\right]\Big|_{r_h}^{\infty}+\pi \beta 
\left[\frac{2^{1/4}5f_1 A_3'\left(A_3'{}^2\right)^{1/4}\delta 
A_3}{\left(\sin{\theta}\right)^{3/2}}\right]\Bigg|_{\theta=0}^{\theta=\pi}\int_{
r_h}^\infty \frac{dr}{w^3}+\delta\mathcal{B}~.
\end{equation}
The condition $\delta \mathcal{I}_E=0$ leads to 
\begin{equation}
    \delta\mathcal{B}=-4\pi \beta \left[ww' \delta F-wF'\delta w+2wF \delta 
w'-2F w^2 \phi' \delta\phi\right]\Big|_{r_h}^{\infty}-\pi \beta 
\left[-2^{1/4}5f_1 Q_m \left(Q_m^2\right)^{1/4}\delta 
A_3\right]\Big|_{\theta=0}^{\theta=\pi}\left[\frac{r}{\alpha^2 
\sqrt{\alpha^2+r^2}}\right]\Bigg|_{r_h}^{\infty}~, \label{deltaB}
\end{equation}
where, in the last term, we have made use of the solutions (\ref{A3}) and 
(\ref{w}). The first contribution in $\delta\mathcal{B}$ originates from the 
gravitational and scalar fields, while the second one arises from the 
non-linear electromagnetic field and the scalar field. It is important to 
emphasize that, in the limit $\alpha \to 0$ ($w\to r$), the last boundary term 
at spatial infinity vanishes, however the difference $\left[\frac{r}{\alpha^2 
\sqrt{\alpha^2+r^2}}\right]\Big|_{r_h}^{\infty}$ approaches $\frac{1}{2r_h^2}$, 
which precisely reproduces the standard result obtained in the usual case 
$w=r$. By contrast, the presence of the phantom scalar field renders this 
contribution nonzero. Using the relation (\ref{Qm2}), the boundary variation 
further reduces to
\begin{equation}
    \delta\mathcal{B}=-4\pi \beta \left[w w' \delta F-wF'\delta w+2wF \delta 
w'-2F w^2 \phi' \delta\phi\right]\Big|_{r_h}^{\infty}-15\pi \beta h_1 \alpha^4 
\delta \alpha\left[\frac{r}{\alpha^2 
\sqrt{\alpha^2+r^2}}\right]\Bigg|_{r_h}^{\infty}~, \label{boundary}
\end{equation}
where the parameter $\alpha$ is treated as an integration constant that is not 
fixed by the theory and is therefore allowed to vary.

We now fix the variation of the boundary term $\delta \mathcal{B}$ such that 
the above relation is satisfied. At spatial infinity, the variation of the 
relevant functions yields
\begin{gather}
\delta F =-\frac{15 \alpha ^2 \delta \alpha h_1}{4 r} + 
\mathcal{O}(r^{-3})~,\hspace{1.5em}\delta w =\frac{\alpha \delta \alpha}{r} + 
\mathcal{O}(r^{-3})~,\hspace{1.5em}
\delta \phi= -\frac{\delta \alpha}{r} + \mathcal{O}(r^{-3})~,
\end{gather}
while at the event horizon of the black hole we have the following relations 
\begin{gather}
 \delta F|_{r_h} = -F'(r_h)\delta r_h~,\hspace{1.5em}\delta w|_{r_h} =\delta 
w(r_h) - w'(r_h)\delta r_h~,\hspace{1.5em}
\delta \phi|_{r_h}= \delta \phi(r_h) - \phi'(r_h) \delta r_h~,
\end{gather}
where the horizon condition $F(r_h)=0$ has been utilized. The variation 
$\delta\mathcal{B}$ contains two contributions, one from spatial infinity and 
one from the event horizon of the black hole, namely $\delta 
\mathcal{B}|_{r_h}^{\infty} = \delta \mathcal{B}(\infty) - \delta 
\mathcal{B}(r_h)$. In particular, from the evaluation of (\ref{boundary}) at 
spatial infinity, we observe that the only finite contribution from the 
gravity-scalar sector arises from the first term, $ww'\delta F$. All remaining 
terms fall off sufficiently fast at large distances and therefore do not 
contribute at spatial infinity. 

Taking into account that the NLED boundary term also yields a finite 
contribution at spatial infinity, we find 
\begin{equation}
    \delta \mathcal{B}(\infty) - 15 \pi  \alpha ^2 \beta  \delta \alpha h_1+ 15 
\pi  \alpha ^2 \beta  \delta \alpha h_1= 0~ \to \delta \mathcal{B}(\infty) =0 
\to \mathcal{B}(\infty) =C_1~,
\end{equation}
where $C_1$ is an integration constant satisfying $\delta C_1=0$ and is 
independent of the boundary values of the fields.
At the event horizon, we obtain
\begin{equation}
    -\delta \mathcal{B}(r_h) + 2\pi \beta F'(r_h)\delta w^2(r_h) -\frac{15 \pi  
\alpha ^2 \beta  \delta \alpha h_1 r_h}{\sqrt{\alpha ^2+r_h^2}}=0~.
\end{equation}
The last term, proportional to $\delta \alpha$, can be rewritten in terms of 
$\delta Q_m$ by means of relation~(\ref{Qm2}), yielding
\begin{equation}
    -\delta \mathcal{B}(r_h) + 2\pi \delta A(r_h) \mp \frac{15\pi \beta 
h_1}{2^{1/4}}\left(\frac{2f_1}{3h_1}\right)^{3/5}\sqrt{|Q_m|}\frac{r_h}{\sqrt{
\alpha ^2+r_h^2}} \delta Q_m=0~,
\end{equation}
where $\mp$ corresponds to $Q_m=\pm |Q_m|$. Through (\ref{Qm2}) $\alpha$ and 
$Q_m$ are related, however keeping both these constants at this stage leads to 
a simplified expression and the relation with the chemical potential associated 
with the magnetic charge in the following arguments is more natural.  
We now adopt the grand canonical ensemble, in which the black hole is allowed 
to exchange magnetic charge with its environment, while the temperature and the 
chemical potential associated with the magnetic charge are kept fixed. This 
implies that, upon identifying 
\begin{equation}
    \Phi=\mp \frac{15\pi 
h_1}{2^{1/4}}\left(\frac{2f_1}{3h_1}\right)^{3/5}\sqrt{|Q_m|}\frac{r_h}{\sqrt{
\alpha ^2+r_h^2}}~,
\end{equation}
the variation of the boundary term at the horizon reduces to 
\begin{equation}
    \delta\mathcal{B}(r_h) = 2\pi \delta A(r_h) + \beta \Phi \delta Q_m~.
\end{equation}
Here $A=4\pi w^2(r_h)$ denotes the horizon area of the black hole. Therefore, 
in the grand canonical ensemble, the boundary term can be integrated to give
\begin{equation}
    \mathcal{B}(r_h) = 2\pi A(r_h) + \beta\Phi Q_m + C_2~,
\end{equation}
where $C_2$ is another integration constant satisfying $\delta C_2=0$ and is 
independent of the boundary values of the fields.
 As a result, the Euclidean action takes the form 
\begin{equation}
    \mathcal{I}_E \equiv \beta \mathcal{G} \equiv \beta\left(\mathcal{M} - 
T\mathcal{S} - \Phi Q_m\right) = \mathcal{B}= C_{12} - 2\pi A(r_h) - \beta\Phi 
Q_m~, \label{euclidean}
\end{equation}
where $C_{12} = C_1-C_2$. Imposing that the free energy vanishes for $\alpha=0$ 
(which corresponds to $r_h=0= Q_m)$, namely the absence of a black hole, we 
obtain $C_{12}=0$. Regardless, the value of $C_{12}$ affects only the value of the free energy when the black hole vanishes and not the thermodynamic mass, neither the entropy or magnetic potential of the black hole

Using the standard thermodynamic relations, the conserved mass, magnetic 
charge, and entropy of the black hole are then given by
\begin{eqnarray}
    &&\mathcal{M} = \frac{\partial \mathcal{I}_E}{\partial \beta} - 
\beta^{-1}\Phi \frac{\partial \mathcal{I}_E}{\partial \Phi} = 0~, 
\label{fund1}\\
    &&\mathcal{Q} = -\beta^{-1}\frac{\partial \mathcal{I}_E}{\partial \Phi} = 
Q_m~,\\
    &&\mathcal{S} = \left(\beta \frac{\partial}{\partial 
\beta}-1\right)\mathcal{I}_E = 2\pi A(r_h)~. \label{fund3}
\end{eqnarray}
Note that  the thermodynamic mass of the SV spacetime 
vanishes identically, even though the geometry possesses a non-zero geometric 
mass parameter determined from the asymptotic fall-off of the metric.
Additionally, the above  quantities satisfy the first law of black hole 
thermodynamics in the form 
\begin{equation}
    \delta \mathcal{M} \equiv 0 \equiv T \delta \mathcal{S} + \Phi \delta 
\mathcal{Q}~. \label{firstlaw}
\end{equation}
Expressed in terms of the free parameter $\alpha$, the relevant thermodynamic 
quantities read
\begin{eqnarray}
    &&\mathcal{S} = \frac{25}{2} \pi ^2 \alpha ^6 h_1^2~,\\
    &&T \equiv \frac{1}{\beta} = \frac{\sqrt{25 \alpha ^4 h_1^2-16}}{25 \pi  
\alpha ^5 h_1^2}~, \label{temp_hairy}\\
    &&\Phi = -\frac{3^{3/5} \pi  f_1^{2/5} \sqrt{25 \alpha ^4 
h_1^2-16}}{2^{1/10} h_1^{2/5}\alpha}~. 
\end{eqnarray}
As a consistency check, one can verify that substituting these expressions into 
the first law~(\ref{firstlaw}), together with the constraint~(\ref{Qm2}), 
yields an identity upon varying $\alpha$. In particular we have that 
\begin{equation}
    T\delta \mathcal{S} = 3 \pi  \delta\alpha  \sqrt{25 \alpha ^4 h_1^2-16} = - 
\Phi \delta \mathcal{Q}~.
\end{equation}

Moreover, these thermodynamic quantities satisfy a Smarr-type relation, 
\begin{equation}
    T\mathcal{S} +\frac{1}{3}\Phi \mathcal{Q}=0~.
\end{equation}
The free energy of the solution acquires a very simple form and is given by 
\begin{equation}
    \mathcal{G} = -TS - \Phi Q_m = \pi  \alpha  \sqrt{25 \alpha ^4 h_1^2-16} = 
4\pi r_h~,
\end{equation}
which grows linearly with the horizon and therefore is reminiscent of the free 
energy of the Schwarzschild black hole solution \cite{Altamirano:2014tva}.

The vanishing of the thermodynamic mass $\mathcal{M}=0$ does not imply the 
absence of gravitational interaction, nor does it indicate that the spacetime 
is trivial. Rather, it shows that the conserved charge associated with time 
translations does not receive a net contribution from the boundary structure of 
the theory. Physically, the geometry still contains a horizon with finite 
temperature and non-vanishing entropy, indicating that the black hole stores 
microscopic degrees of freedom. The gravitational and matter contributions 
precisely balance so that the asymptotic charge vanishes, even though the local 
geometry remains non-trivial. From a thermodynamic perspective, the first law
$\delta \mathcal{M}=T\delta\mathcal{S}+\Phi\delta\mathcal{Q}=0$
implies that variations of the entropy are entirely determined by variations of 
the magnetic charge. The Smarr-type relation
$T\mathcal{S}+\frac13\Phi\mathcal{Q}=0$
further confirms that the scaling balance between the horizon degrees of 
freedom and the electromagnetic sector leaves no room for an independent 
extensive energy contribution. It is also worth noting that, although this is evident from the calculations, the grand-canonical ensemble applies here only in a constrained sense, since the solution space is one-dimensional.

From the geometric perspective, the metric function asymptotically approaches 
the Schwarzschild form with a $\mathcal{O}(r^{-1})$ fall-off at large 
distances. Nevertheless, semiclassical arguments based on the Euclidean action 
show that the spacetime is thermodynamically massless and that the Komar mass 
$M_{\text{Komar}}$ cannot be identified with the thermodynamic mass 
$\mathcal{M}$. The latter encodes the physical energy relevant for black-hole 
thermodynamics, and in the present theory it vanishes due to the specific 
structure of the matter couplings. More generally, the entropy is determined by 
the theory and the solution, while the temperature is fixed purely by the 
spacetime geometry through the elimination of conical singularities at the 
event horizon in the Euclidean section. It is important to emphasize that, without constraining the field equations by imposing $w(r) = \sqrt{r^2+\alpha^2}$, neither a reduction of the parameter space nor the vanishing of the thermodynamic mass would occur. In fact, as we show in the next subsection, for $w(r) = r$ and a constant scalar field, the parameter space remains unchanged. Our computation therefore illustrates 
that in modified gravity or NLED theories the notion of mass is not merely a 
geometric integration constant, but a quantity determined by the full action 
principle.

 Related phenomena have appeared in several contexts in the literature. In 
\cite{Martinez:2006an}, where a self-interacting scalar field coupled to linear 
electrodynamics was considered, massless AdS black holes were obtained due to 
the back-reaction of the scalar field on the geometry, which modifies the 
asymptotic fall-off relative to standard AdS solutions. In the context of 
stealth black holes, shifts in the black-hole mass have also been reported. 
Such shifts may be dictated by the underlying theory \cite{Bakopoulos:2024zke} 
or may remain unconstrained by it \cite{Erices:2024iah}, a situation that 
resembles the SV spacetime when it arises as a solution of the present theory. 
In stealth configurations, computing the mass with or without taking into 
account the specific theory can lead to conflicting results, although only one 
of these definitions is consistent with the first law of thermodynamics and 
with the existence of a well-defined variational principle.

Similar situations have also been discussed in asymptotically Lifshitz 
spacetimes \cite{Liu:2014dva, Bravo-Gaete:2015xea, Fan:2014ala}. In such cases, 
as in the present work, a naive integration of $T\delta S$ with respect to the 
horizon cannot reliably determine the mass, and more formal methods are 
required. Massless and entropy-less black holes were recently discussed in 
\cite{Guajardo:2024hrl}, where the Euclidean action vanishes on-shell and 
reaches an extremum without the need for a boundary term $\mathcal{B}$ (i.e.\ 
$\mathcal{B}=0$). The vanishing of the boundary term has also appeared in the 
context of Lovelock black holes non-minimally coupled to scalar fields 
\cite{Correa:2013bza}. This situation resembles certain wormhole 
configurations. 
For instance, in the Ellis wormhole \cite{Ellis:1973yv}, to which our theory 
reduces when $h_1=0$, both the action and the boundary term vanish on-shell. 
However, in the cases discussed in \cite{Guajardo:2024hrl, Correa:2013bza} the 
solutions still possess an event horizon and finite temperature due to the 
conical singularity structure in the Euclidean section, and therefore 
correspond 
to genuine black holes.

\section{Scalar-free configuration}
\label{Scalarfree}

Having analyzed the thermodynamic properties of the SV spacetime supported by 
both the nonlinear electrodynamics and the scalar sector, we proceed to 
examine whether the same theory also admits black-hole configurations in which 
the scalar field does not participate dynamically. Such a configuration allows 
us to isolate the role of the scalar sector in the regularization mechanism and 
to compare the resulting thermodynamic behavior with that of the SV geometry 
obtained in the previous sections.

In the   theory characterized by action (\ref{generalaction}), with 
(\ref{potential}) and (\ref{LF}), 
one may therefore consider a solution in which the scalar field remains 
constant. In order to investigate this possibility we adopt a different metric 
ansatz, namely
\begin{equation}
    ds^2 = -H(\rho) dt^2 + \frac{d\rho^2}{H(\rho)} + \rho^2 d\Omega^2~,
\end{equation}
which corresponds to removing the core scale $\alpha$ introduced in the 
previous subsections while keeping the magnetic potential 
$A_3(\theta) = q \cos{\theta}$. In this case the field equations admit the 
solution
\begin{eqnarray}
    &&\phi(\rho)=\text{const}=\phi_0~,\\
    &&H(\rho) = 1-\frac{2 \mu }{\rho }+\frac{f_1 
\left(q^2\right)^{5/4}}{2^{3/4} \rho ^3}-\frac{h_1 \rho ^2}{6}\cos^5{\phi_0}~,\\
    &&h_1 \cos{\phi_0}\sin{\phi_0}=0~.
\end{eqnarray}

For $\phi_0=0$, the solution describes an asymptotically AdS black hole, 
\begin{equation}
    H(\rho) = 1-\frac{2 \mu }{\rho }+\frac{f_1 \left(q^2\right)^{5/4}}{2^{3/4} 
\rho ^3}-\frac{h_1 \rho ^2}{6}~.
\end{equation}

On the other hand, for $\phi_0=\frac{\pi}{2}$ the spacetime exhibits the same 
asymptotic behaviour as the solution discussed in the previous subsections, 
namely 
\begin{equation}
    H(\rho) = 1-\frac{2 \mu }{\rho }+\frac{f_1 \left(q^2\right)^{5/4}}{2^{3/4} 
\rho ^3}~.
\end{equation}
In this latter case, the thermodynamic properties of this configuration can be 
compared with those of the SV spacetime in order to determine which geometry is 
thermodynamically preferred. In equilibrium thermodynamics the configuration 
favoured by nature is the one with the lowest free energy within the same phase 
space. Therefore, the SV and scalar-free black holes must be placed in the same 
heat bath, characterized by the same temperature and magnetic chemical 
potential. Once these conditions are imposed, the configuration with the 
least free energy represents the thermodynamically stable state.

The thermodynamic analysis follows a similar procedure to the one presented in 
the previous section, although the form of the boundary contribution differs 
from Eq.~(\ref{deltaB}). In particular, the variation of the boundary term now 
reads 
\begin{equation}
    \delta\mathcal{B}=-4\pi \beta_0 \left[\rho~\delta 
H\right]\Big|_{\rho_h}^{\infty}-2^{5/4}5\pi \beta_0 f_1 q 
\left(q^2\right)^{1/4}\delta 
q\left[-\frac{1}{2\rho^2}\right]\Bigg|_{\rho_h}^{\infty}~, 
\end{equation}
where $\beta_0=4\pi/H'(\rho_h)$ defines the Hawking temperature (we use the 
index $0$ to denote the $\phi = \phi_0$ case)
\begin{equation}
    T_0\equiv \frac{1}{\beta_0}=\frac{\rho_h^3-2^{1/4} f_1 
\left(q^2\right)^{5/4}}{4 \pi  \rho_h^4}~,
\end{equation}
and the variations of the boundary term become
\begin{eqnarray}
    &&\delta\mathcal{B}\left(\infty\right)=8\pi \beta_0~\delta \mu~, \quad 
\quad \quad \quad \quad \quad \ \ \mathcal{B}\left(\infty\right)=8\pi 
\beta_0\mu~,\\
    &&\delta\mathcal{B}\left(\rho_h\right)=2\pi \delta 
A_0\left(\rho_h\right)+\beta_0 \Phi_0 \delta q~, \quad 
\mathcal{B}\left(\rho_h\right)=2\pi A_0\left(\rho_h\right)+\beta_0 \Phi_0 q~,
\end{eqnarray}
with $\Phi_0=2^{1/4} 5\pi f_1 q \left(q^2\right)^{1/4}/\rho_h^2$, where we have 
invoked the grand canonical ensemble, assuming constant temperature and 
constant magnetic chemical potential.

Therefore, up to a constant without variation, we have for the Euclidean action 
and the free energy of the scalar free black hole
\begin{eqnarray}
    &&\mathcal{I}_E^0 \equiv \beta_0 \mathcal{G}_0 \equiv 
\beta_0\left(\mathcal{M}_0 - T_0\mathcal{S}_0 - \Phi_0 q\right) = 
\mathcal{B}=8\pi \beta_0\mu-2\pi A_0\left(\rho_h\right)-\beta_0 \Phi_0 q~, \\
    &&\mathcal{G}_0=8\pi\mu-\frac{2\pi}{\beta_0}A_0\left(\rho_h\right)-\Phi_0 
q~,
\end{eqnarray}
and
\begin{eqnarray}
    &&\mathcal{M}_0 = \frac{\partial \mathcal{I}_E^0}{\partial \beta_0} - 
\beta_0^{-1}\Phi_0 \frac{\partial \mathcal{I}_E^0}{\partial \Phi_0} = 8\pi\mu~, 
\\
    &&\mathcal{Q}_0 = -\beta_0^{-1}\frac{\partial \mathcal{I}_E^0}{\partial 
\Phi_0} =q~,\\
    &&\mathcal{S}_0 = \left(\beta_0 \frac{\partial}{\partial 
\beta_0}-1\right)\mathcal{I}_E^0 = 2\pi A_0(\rho_h)=8\pi^2\rho_h^2~. 
\end{eqnarray}
 One then can verify that the aforementioned relations satisfy the standard 
first law of thermodynamics
\begin{equation}
    \delta \mathcal{M}_0 =  T_0  \delta \mathcal{S}_0 + \Phi_0 \delta 
\mathcal{Q}_0~.
\end{equation}

\begin{figure}[h]
\centering
  \includegraphics[width=0.65\textwidth]{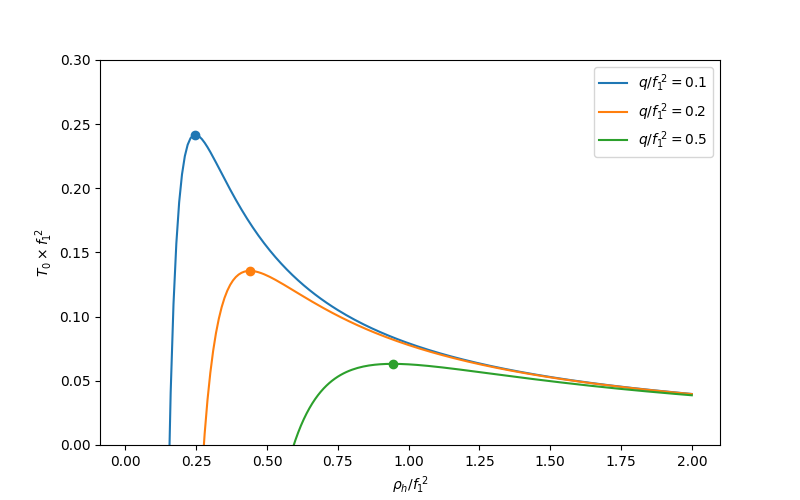}
\caption{{\it{Temperature $T_0(\rho_h)$ of the scalar-free black-hole solution 
as a function of the horizon radius $\rho_h$ for several values of the 
dimensionless parameter $q/f_1^{\,2}$. The dots mark the extrema of the 
temperature, which separate the small and large black-hole branches and signal 
the onset of thermodynamic phase transitions.}}}
\label{Temp_scalar_free}
\end{figure}

In FIG.~\ref{Temp_scalar_free} we present the temperature of the scalar-free 
solution. As can be seen from the figure, its behavior closely resembles that 
of the Reissner–Nordström case: the temperature admits a maximum corresponding 
to a transition between large and small black-hole branches and eventually 
approaches extremality where $T=0$. For a given temperature two distinct black 
holes may therefore exist, with the large branch being thermodynamically 
unstable and the small branch stable.

For a given temperature and magnetic chemical potential in the same theory 
(with identical values of $h_1$ and $f_1$), the conditions
\begin{equation}
    T\left(\alpha\right)=T_0\left(q,\rho_h\right)~,\quad 
\Phi\left(\alpha\right)=\Phi_0\left(q,\rho_h\right)~,
\end{equation}
determine the parameters $q$ and $\rho_h$ as functions of $\alpha$. The first 
relation yields
\begin{equation}
    q=\pm \frac{\rho_h^{6/5}}{2^{1/10} 5^{4/5}f_1^{2/5}} \left(25-\frac{4 
\rho_h \sqrt{25 \alpha ^4 h_1^2-16}}{\alpha ^5 h_1^2}\right)^{2/5}~, 
\label{relation_q}
\end{equation}
while the second relation requires selecting the negative branch of $q$, which 
leads to 
\begin{equation}
    2^{1/5} \alpha h_1^{2/5} \left(25-\frac{4 \rho_h \sqrt{25 \alpha ^4 
h_1^2-16}}{\alpha ^5 h_1^2}\right)^{3/5}=5^{1/5} 3^{3/5} \rho_h^{1/5} \sqrt{25 
\alpha ^4 h_1^2-16}~.
\end{equation}

\begin{figure}[h]
\centering
  \includegraphics[width=0.45\textwidth]{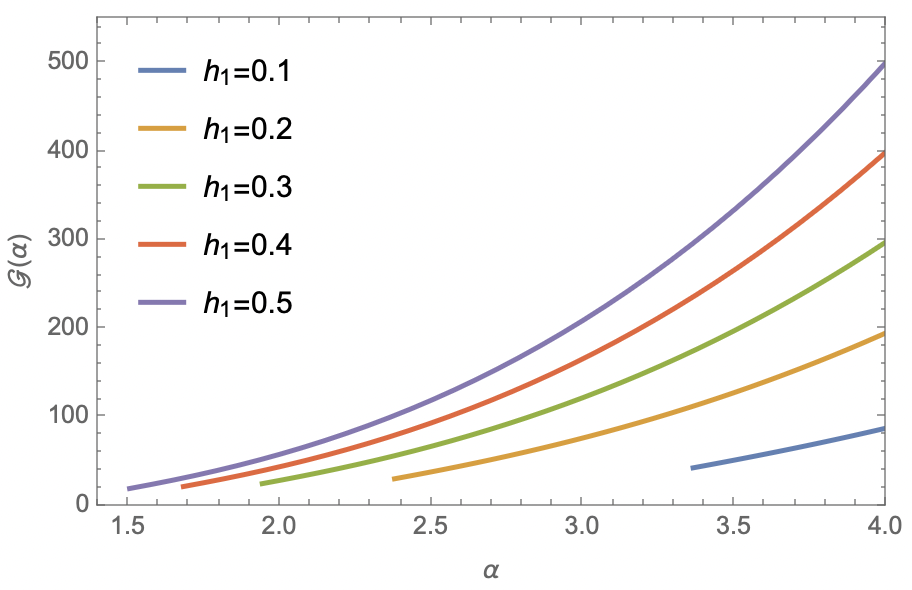}
  \includegraphics[width=0.45\textwidth]{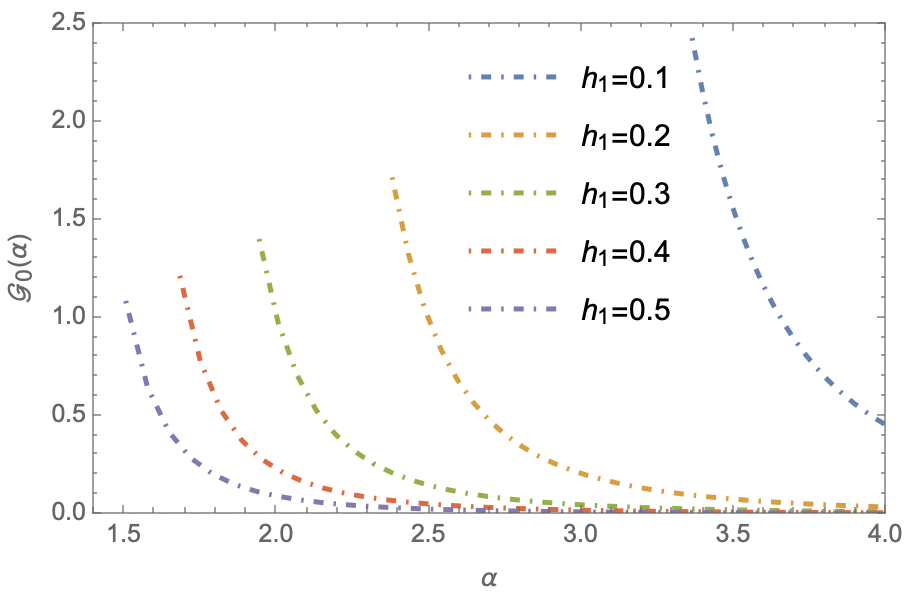}\\
  \includegraphics[width=0.45\textwidth]{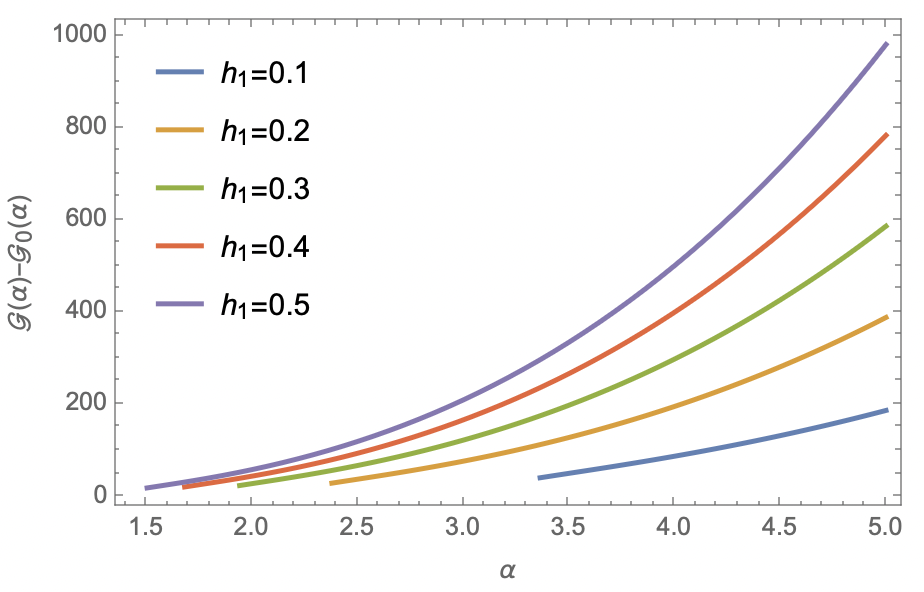}
\caption{{\it{Free energy of the two configurations as a function of the core 
scale $\alpha$ for several values of $h_1$. The upper-left panel shows the free 
energy $\mathcal{G}(\alpha)$ of the reconstructed Simpson-Visser regular black 
hole, while the upper-right panel shows the free energy $\mathcal{G}_0(\alpha)$ 
of the corresponding scalar-free black hole in the same heat bath with 
identical temperature and magnetic chemical potential. The lower panel displays 
their difference $\mathcal{G}(\alpha)-\mathcal{G}_0(\alpha)$, which is always 
positive, indicating that the scalar-free configuration is thermodynamically 
preferred.}}}
\label{Fig_free_energy}
\end{figure}

From this relation, one can solve $\rho_h$ for a given value of $\alpha$, 
independently of $f_1$. Remarkably, although the expression for $q$ explicitly 
depends on $f_1$, this dependence cancels out when computing the mass parameter 
$\mu$ from the condition $H(\rho_h)=0$ with relation (\ref{relation_q}),
\begin{equation}
    \mu=\frac{1}{100} \rho_h \left(75-\frac{4 \rho_h \sqrt{25 \alpha ^4 
h_1^2-16}}{\alpha ^5 h_1^2}\right)~.
\end{equation}
In the same way, the expressions for $\beta_0$ and the free energy 
$\mathcal{G}_0$ are also independent of $f_1$,
\begin{eqnarray}
    &&\beta_0=\frac{25 \pi  \alpha ^5 h_1^2}{\sqrt{25 \alpha ^4 h_1^2-16}}~,\\
    &&\mathcal{G}_0=\pi  \left(\frac{4 \rho_h^2 \sqrt{25 \alpha ^4 
h_1^2-16}}{25 \alpha ^5 h_1^2}+\rho_h\right),
\end{eqnarray}
where $\beta_0=\beta$ in (\ref{temp_hairy}) as expected.
Therefore, the free energy used for comparison, $\mathcal{G}_0(\alpha)$, is 
also independent of $f_1$. In FIG.~\ref{Fig_free_energy}, we plot 
$\mathcal{G}(\alpha)$, $\mathcal{G}_0(\alpha)$ and their difference for several 
values of $h_1$. From the figure, we observe that $\mathcal{G}$ increases 
monotonically with $\alpha$, whereas $\mathcal{G}_0(\alpha)$ decreases 
monotonically. Moreover, as $h_1$ increases, $\mathcal{G}$ increases while 
$\mathcal{G}_0$ decreases. 

Most importantly, the free energy of the regular SV black hole is always 
significantly larger than that of the corresponding scalar-free configuration. 
This indicates that the scalar-free solution is thermodynamically preferred 
within the same ensemble. Consequently, the regular black hole characterized by 
a nonvanishing core-scale parameter $\alpha$ is expected to decay towards the 
singular configuration with $\alpha=0$ and a constant scalar field. In this 
sense, when the SV spacetime arises as a solution of the present theory, it 
corresponds to a metastable configuration which is thermodynamically 
disfavoured relative to the scalar-free black hole.

\section{Conclusions}
\label{Conclusions}

In this work we investigated the thermodynamic properties of the Simpson-Visser 
(SV) spacetime. In order to perform a consistent thermodynamic analysis, we 
first reconstructed the theory that supports this spacetime in such a way that 
no integration constant appears in the Lagrangian. In this formulation the 
solution can be expressed in terms of two theory constants and a single free 
integration constant, namely the core scale $\alpha$. In this scenario a 
reduction of charges occurs: while one would normally expect black holes to be 
characterized at least by their geometric mass and magnetic charge, in the 
present case all physical quantities are ultimately determined by the core 
scale. Such a reduction of charges has very recently been shown to be a general 
feature of regular black holes supported by nonlinear electrodynamics (NLED) 
\cite{Bokulic:2025brf}. Asymptotically the spacetime behaves as the 
Schwarzschild black hole, while the point $r=0$ is replaced by a two-sphere of 
radius $\alpha$, rendering the geometry regular. Furthermore, we showed that 
when the theory parameter $h_1$ is set to zero the solution reduces to the 
Ellis wormhole spacetime, supported by a free phantom scalar field. In this 
limit the scalar potential as well as the gauge field vanish.

We then constructed the first law of thermodynamics for the SV spacetime using 
Euclidean methods and the principle of least action. In this approach the 
Euclidean action vanishes on shell by definition, and therefore all 
thermodynamic information is encoded in the boundary term that must be 
introduced in order to ensure a well-defined variational principle. By 
evaluating the variation of this boundary term we obtained the first law of 
black-hole thermodynamics and the free energy of the spacetime. Remarkably, we 
found that at spatial infinity the magnetic contribution arising from the NLED 
sector, together with the scalar contribution, exactly cancels the purely 
gravitational part, leading to a vanishing thermodynamic mass. It is important 
to stress that although this contribution originates from the NLED Lagrangian, 
the existence of the core scale $\alpha$ - which replaces $r^2$ with 
$r^2+\alpha^2$ in the angular part of the metric - is responsible for the 
finiteness of this boundary term. In the absence of the NLED sector the 
cancellation would not occur and the mass would not vanish. This result 
indicates that the first law written a priori in the form 
$\delta\mathcal{M}=T\delta\mathcal{S}+\dots$ and the naive integration of 
$T\delta\mathcal{S}$ cannot in general be trusted to determine the mass of a 
black hole; instead, more formal methods based on the action principle are 
required \cite{Yu:2026syo}. In the literature there exist works (for example 
\cite{Tzikas:2018cvs,Ahmed:2026bwm}) that use the relation 
$\delta\mathcal{M}=T\delta S$ in order to determine either the mass or the 
entropy. This does not imply that these approaches are incorrect; rather, it 
suggests that additional theories beyond GR may 
also admit the same line element as a solution. In such cases logarithmic 
corrections to the entropy may arise naturally, as happens for instance in 
Horndeski gravity \cite{Bakopoulos:2025eps}.

Although the thermodynamic mass vanishes, the spacetime still possesses a 
single event horizon and therefore a well-defined temperature and entropy. The 
temperature is determined by the periodicity of Euclidean time, as expected 
from finite-temperature quantum field theory, while the entropy satisfies the 
standard Bekenstein-Hawking area law since the solution arises within GR with 
minimally coupled fields. Consequently, these configurations represent genuine 
thermodynamic black holes despite the absence of an extensive energy 
contribution. Similar situations have previously appeared in modified gravity 
theories, where black holes may exhibit vanishing thermodynamic mass while 
maintaining nontrivial horizon structure and thermodynamic behaviour. The 
corresponding Smarr relation satisfied by the solution further confirms the 
internal thermodynamic consistency of the system.

Finally, we examined the scalar-free configuration that naturally arises within 
the same theory. In this case an asymptotically flat black-hole solution is 
obtained with properties resembling those of the Reissner-Nordström spacetime. 
This solution obeys the standard first law of thermodynamics, and the 
temperature analysis shows that small black holes can remain in thermal 
equilibrium with a heat bath due to their positive heat capacity. In order to 
compare the two configurations within the same ensemble, we imposed equal 
temperature and magnetic chemical potential for both solutions and related the 
core parameter $\alpha$ of the SV spacetime to the parameters $\rho_h$ and $q$ 
of the scalar-free solution. We then evaluated their free energies and found 
that the scalar-free configuration always possesses the lower free energy. This 
indicates that it is thermodynamically preferred. Consequently, when the SV 
spacetime arises as a solution of the present theory, it is expected to relax 
toward the scalar-free black hole configuration, suggesting that the regular SV 
geometry represents a metastable state within the thermodynamic phase space.

 \begin{acknowledgments}
  ENS acknowledges the contribution of the LISA   CosWG,  and of COST Actions CA21136 ``Addressing observational tensions in 
cosmology with systematics and fundamental physics (CosmoVerse)'', CA21106 
``COSMIC WISPers in the Dark Universe: Theory, astrophysics and experiments'', 
 CA23130 ``Bridging high and low energies in search of quantum gravity 
(BridgeQG)'', and CA24101
 ``Testing Fundamental Physics with Seismology''. Z.-Y. Tang is supported by IBS under the project code IBS-R018-D3.
  
 \end{acknowledgments}

\appendix

\section{A radial magnetic field}
\label{AppendixA}

Considering the $(3+1)$ decomposition of spacetime \cite{Alcubierre:2008}
\begin{equation}
    ds^2=-\alpha^2dt^2+\gamma_{ij}\left(dx^i+\beta^i 
dt\right)\left(dx^j+\beta^j dt\right)~,
\end{equation}
we may define the electric and magnetic fields~\cite{Alcubierre:2009ij}
\begin{equation}
    E^a :=n_b F^{ab}~, \quad B^a:=n_b{}^*F^{ab}~,
\end{equation}
with respect to an Eulerian observer 
$n^a=\left(1/\alpha,-\beta^i/\alpha\right)$ whose worldline is orthogonal to the 
spacelike hypersurfaces. For the static and spherically symmetric spacetime 
described by the metric ansatz (\ref{metric}), the components of the magnetic 
field take the form
\begin{eqnarray}
    B^r&=&\frac{\sqrt{F(r)}}{w(r)^2 \sin{\theta}}F_{\theta\varphi}~,\\
    B^\theta&=&-\frac{\sqrt{F(r)}}{w(r)^2 \sin{\theta}}F_{r\varphi}~,\\
    B^\varphi&=&\frac{\sqrt{F(r)}}{w(r)^2 \sin{\theta}}F_{r\theta}~,
\end{eqnarray}
where a purely radial magnetic field requires $F_{r\varphi}=0$ and 
$F_{r\theta}=0$.

\bibliography{Refs}

@article{Bronnikov:2021uta,
    author = "Bronnikov, Kirill A. and Walia, Rahul Kumar",
    title = "{Field sources for Simpson-Visser spacetimes}",
    eprint = "2112.13198",
    archivePrefix = "arXiv",
    primaryClass = "gr-qc",
    doi = "10.1103/PhysRevD.105.044039",
    journal = "Phys. Rev. D",
    volume = "105",
    number = "4",
    pages = "044039",
    year = "2022"
}

@article{Alcubierre:2008,
    author = "Alcubierre, Miguel",
    title = "{Introduction to $3+1$ Numerical Relativity}",
    journal = "Oxford University Press, New York",
    year = "2009"
}

@article{Alcubierre:2009ij,
    author = "Alcubierre, Miguel and Degollado, Juan Carlos and Salgado, Marcelo",
    title = "{The Einstein-Maxwell system in 3+1 form and initial data for multiple charged black holes}",
    eprint = "0907.1151",
    archivePrefix = "arXiv",
    primaryClass = "gr-qc",
    doi = "10.1103/PhysRevD.80.104022",
    journal = "Phys. Rev. D",
    volume = "80",
    pages = "104022",
    year = "2009"
}

@article{Bakopoulos:2024hah,
    author = "Bakopoulos, Athanasios and Karakasis, Thanasis and Mavromatos, Nick E. and Nakas, Theodoros and Papantonopoulos, Eleftherios",
    title = "{Exact black holes in string-inspired Euler-Heisenberg theory}",
    eprint = "2402.12459",
    archivePrefix = "arXiv",
    primaryClass = "hep-th",
    reportNumber = "KCL-TH-PH/2024-03",
    doi = "10.1103/PhysRevD.110.024014",
    journal = "Phys. Rev. D",
    volume = "110",
    number = "2",
    pages = "024014",
    year = "2024"
}

@article{Karakasis:2023hni,
    author = "Karakasis, Thanasis and Mavromatos, Nick E. and Papantonopoulos, Eleftherios",
    title = "{Regular compact objects with scalar hair}",
    eprint = "2305.00058",
    archivePrefix = "arXiv",
    primaryClass = "gr-qc",
    reportNumber = "KCL-PH-TH/2023-02",
    doi = "10.1103/PhysRevD.108.024001",
    journal = "Phys. Rev. D",
    volume = "108",
    number = "2",
    pages = "024001",
    year = "2023"
}

@article{Gibbons:1976ue,
    author = "Gibbons, G. W. and Hawking, S. W.",
    title = "{Action Integrals and Partition Functions in Quantum Gravity}",
    reportNumber = "PRINT-76-0995 (CAMBRIDGE)",
    doi = "10.1103/PhysRevD.15.2752",
    journal = "Phys. Rev. D",
    volume = "15",
    pages = "2752--2756",
    year = "1977"
}

@article{Alkac:2024hvu,
    author = "Alkac, Gokhan and Guajardo, Luis and Ozsahin, Hikmet",
    title = "{Microscopic entropy of static black holes in 3D Lovelock gravities}",
    eprint = "2409.03865",
    archivePrefix = "arXiv",
    primaryClass = "hep-th",
    doi = "10.1103/PhysRevD.111.044006",
    journal = "Phys. Rev. D",
    volume = "111",
    number = "4",
    pages = "044006",
    year = "2025"
}

@article{Bakopoulos:2025eps,
    author = "Bakopoulos, Athanasios and Karakasis, Thanasis",
    title = "{Thermodynamic analysis of shift-symmetric black-hole spacetimes in Horndeski gravity}",
    eprint = "2502.07919",
    archivePrefix = "arXiv",
    primaryClass = "gr-qc",
    month = "2",
    year = "2025"
}

@article{Bakopoulos:2024zke,
    author = "Bakopoulos, Athanasios and Karakasis, Thanasis and Papantonopoulos, Eleftherios",
    title = "{Thermodynamics of stealth black holes}",
    eprint = "2410.14451",
    archivePrefix = "arXiv",
    primaryClass = "hep-th",
    doi = "10.1103/PhysRevD.111.024065",
    journal = "Phys. Rev. D",
    volume = "111",
    number = "2",
    pages = "024065",
    year = "2025"
}

@article{Bakopoulos:2024ogt,
    author = "Bakopoulos, Athanasios and Chatzifotis, Nikos and Karakasis, Thanasis",
    title = "{Thermodynamics of black holes featuring primary scalar hair}",
    eprint = "2404.07522",
    archivePrefix = "arXiv",
    primaryClass = "hep-th",
    doi = "10.1103/PhysRevD.110.L101502",
    journal = "Phys. Rev. D",
    volume = "110",
    number = "10",
    pages = "L101502",
    year = "2024"
}

@article{Simpson:2018tsi,
    author = "Simpson, Alex and Visser, Matt",
    title = "{Black-bounce to traversable wormhole}",
    eprint = "1812.07114",
    archivePrefix = "arXiv",
    primaryClass = "gr-qc",
    doi = "10.1088/1475-7516/2019/02/042",
    journal = "JCAP",
    volume = "02",
    pages = "042",
    year = "2019"
}

@article{Simpson:2019cer,
    author = "Simpson, Alex and Martin-Moruno, Prado and Visser, Matt",
    title = "{Vaidya spacetimes, black-bounces, and traversable wormholes}",
    eprint = "1902.04232",
    archivePrefix = "arXiv",
    primaryClass = "gr-qc",
    doi = "10.1088/1361-6382/ab28a5",
    journal = "Class. Quant. Grav.",
    volume = "36",
    number = "14",
    pages = "145007",
    year = "2019"
}

@article{Lobo:2020ffi,
    author = "Lobo, Francisco S. N. and Rodrigues, Manuel E. and de Sousa Silva, Marcos V. and Simpson, Alex and Visser, Matt",
    title = "{Novel black-bounce spacetimes: wormholes, regularity, energy conditions, and causal structure}",
    eprint = "2009.12057",
    archivePrefix = "arXiv",
    primaryClass = "gr-qc",
    doi = "10.1103/PhysRevD.103.084052",
    journal = "Phys. Rev. D",
    volume = "103",
    number = "8",
    pages = "084052",
    year = "2021"
}

@article{Franzin:2021vnj,
    author = "Franzin, Edgardo and Liberati, Stefano and Mazza, Jacopo and Simpson, Alex and Visser, Matt",
    title = "{Charged black-bounce spacetimes}",
    eprint = "2104.11376",
    archivePrefix = "arXiv",
    primaryClass = "gr-qc",
    doi = "10.1088/1475-7516/2021/07/036",
    journal = "JCAP",
    volume = "07",
    pages = "036",
    year = "2021"
}

@article{Mazza:2021rgq,
    author = "Mazza, Jacopo and Franzin, Edgardo and Liberati, Stefano",
    title = "{A novel family of rotating black hole mimickers}",
    eprint = "2102.01105",
    archivePrefix = "arXiv",
    primaryClass = "gr-qc",
    doi = "10.1088/1475-7516/2021/04/082",
    journal = "JCAP",
    volume = "04",
    pages = "082",
    year = "2021"
}

@article{Shaikh:2021yux,
    author = "Shaikh, Rajibul and Pal, Kunal and Pal, Kuntal and Sarkar, Tapobrata",
    title = "{Constraining alternatives to the Kerr black hole}",
    eprint = "2102.04299",
    archivePrefix = "arXiv",
    primaryClass = "gr-qc",
    doi = "10.1093/mnras/stab1779",
    journal = "Mon. Not. Roy. Astron. Soc.",
    volume = "506",
    number = "1",
    pages = "1229--1236",
    year = "2021"
}

@article{Guo:2021wid,
    author = "Guo, Yang and Miao, Yan-Gang",
    title = "{Charged black-bounce spacetimes: Photon rings, shadows and observational appearances}",
    eprint = "2112.01747",
    archivePrefix = "arXiv",
    primaryClass = "gr-qc",
    doi = "10.1016/j.nuclphysb.2022.115938",
    journal = "Nucl. Phys. B",
    volume = "983",
    pages = "115938",
    year = "2022"
}

@article{Bronnikov:2005gm,
    author = "Bronnikov, K. A. and Fabris, J. C.",
    title = "{Regular phantom black holes}",
    eprint = "gr-qc/0511109",
    archivePrefix = "arXiv",
    doi = "10.1103/PhysRevLett.96.251101",
    journal = "Phys. Rev. Lett.",
    volume = "96",
    pages = "251101",
    year = "2006"
}

@article{Ayon-Beato:2000mjt,
    author = "Ayon-Beato, Eloy and Garcia, Alberto",
    title = "{The Bardeen model as a nonlinear magnetic monopole}",
    eprint = "gr-qc/0009077",
    archivePrefix = "arXiv",
    doi = "10.1016/S0370-2693(00)01125-4",
    journal = "Phys. Lett. B",
    volume = "493",
    pages = "149--152",
    year = "2000"
}

@article{Lan:2023cvz,
    author = "Lan, Chen and Yang, Hao and Guo, Yang and Miao, Yan-Gang",
    title = "{Regular Black Holes: A Short Topic Review}",
    eprint = "2303.11696",
    archivePrefix = "arXiv",
    primaryClass = "gr-qc",
    doi = "10.1007/s10773-023-05454-1",
    journal = "Int. J. Theor. Phys.",
    volume = "62",
    number = "9",
    pages = "202",
    year = "2023"
}

@article{Tzikas:2018cvs,
    author = "Tzikas, Athanasios G.",
    title = "{Bardeen black hole chemistry}",
    eprint = "1811.01104",
    archivePrefix = "arXiv",
    primaryClass = "gr-qc",
    doi = "10.1016/j.physletb.2018.11.036",
    journal = "Phys. Lett. B",
    volume = "788",
    pages = "219--224",
    year = "2019"
}

@article{Fan:2016hvf,
    author = "Fan, Zhong-Ying and Wang, Xiaobao",
    title = "{Construction of Regular Black Holes in General Relativity}",
    eprint = "1610.02636",
    archivePrefix = "arXiv",
    primaryClass = "gr-qc",
    doi = "10.1103/PhysRevD.94.124027",
    journal = "Phys. Rev. D",
    volume = "94",
    number = "12",
    pages = "124027",
    year = "2016"
}

@article{Ellis:1973yv,
    author = "Ellis, H. G.",
    title = "{Ether flow through a drainhole - a particle model in general relativity}",
    doi = "10.1063/1.1666161",
    journal = "J. Math. Phys.",
    volume = "14",
    pages = "104--118",
    year = "1973"
}

@article{Karakasis:2021tqx,
    author = "Karakasis, Thanasis and Papantonopoulos, Eleftherios and Vlachos, Christoforos",
    title = "{f(R) gravity wormholes sourced by a phantom scalar field}",
    eprint = "2107.09713",
    archivePrefix = "arXiv",
    primaryClass = "gr-qc",
    doi = "10.1103/PhysRevD.105.024006",
    journal = "Phys. Rev. D",
    volume = "105",
    number = "2",
    pages = "024006",
    year = "2022"
}

@article{Bakopoulos:2023tso,
    author = "Bakopoulos, Athanasios and Chatzifotis, Nikos and Erices, Cristian and Papantonopoulos, Eleftherios",
    title = "{Stealth Ellis wormholes in Horndeski theories}",
    eprint = "2306.16768",
    archivePrefix = "arXiv",
    primaryClass = "hep-th",
    doi = "10.1088/1475-7516/2023/11/055",
    journal = "JCAP",
    volume = "11",
    pages = "055",
    year = "2023"
}

@article{Martinez:2006an,
    author = "Martinez, Cristian and Troncoso, Ricardo",
    title = "{Electrically charged black hole with scalar hair}",
    eprint = "hep-th/0606130",
    archivePrefix = "arXiv",
    reportNumber = "CECS-PHY-06-11",
    doi = "10.1103/PhysRevD.74.064007",
    journal = "Phys. Rev. D",
    volume = "74",
    pages = "064007",
    year = "2006"
}

@article{Erices:2024iah,
    author = "Erices, Cristi{\'a}n and Guajardo, Luis and Lara, Kristiansen",
    title = "{Reverse stealth construction and its thermodynamic imprints}",
    eprint = "2410.13719",
    archivePrefix = "arXiv",
    primaryClass = "gr-qc",
    doi = "10.1088/1475-7516/2025/03/051",
    journal = "JCAP",
    volume = "03",
    pages = "051",
    year = "2025"
}

@article{Liu:2014dva,
    author = {Liu, Hai-Shan and L{\"u}, H.},
    title = "{Thermodynamics of Lifshitz Black Holes}",
    eprint = "1410.6181",
    archivePrefix = "arXiv",
    primaryClass = "hep-th",
    doi = "10.1007/JHEP12(2014)071",
    journal = "JHEP",
    volume = "12",
    pages = "071",
    year = "2014"
}

@article{Bravo-Gaete:2015xea,
    author = "Bravo-Gaete, Moises and Hassaine, Mokhtar",
    title = "{Thermodynamics of charged Lifshitz black holes with quadratic corrections}",
    eprint = "1501.03348",
    archivePrefix = "arXiv",
    primaryClass = "hep-th",
    doi = "10.1103/PhysRevD.91.064038",
    journal = "Phys. Rev. D",
    volume = "91",
    number = "6",
    pages = "064038",
    year = "2015"
}

@article{Fan:2014ala,
    author = "Fan, Zhong-Ying and Lu, H.",
    title = "{Thermodynamical First Laws of Black Holes in Quadratically-Extended Gravities}",
    eprint = "1501.00006",
    archivePrefix = "arXiv",
    primaryClass = "hep-th",
    doi = "10.1103/PhysRevD.91.064009",
    journal = "Phys. Rev. D",
    volume = "91",
    number = "6",
    pages = "064009",
    year = "2015"
}

@article{Guajardo:2024hrl,
    author = "Guajardo, Luis and Oliva, Julio",
    title = "{Primary scalar hair in Gauss{\textendash}Bonnet black holes with Thurston horizons}",
    eprint = "2412.20134",
    archivePrefix = "arXiv",
    primaryClass = "hep-th",
    doi = "10.1140/epjc/s10052-025-13869-9",
    journal = "Eur. Phys. J. C",
    volume = "85",
    number = "2",
    pages = "139",
    year = "2025"
}

@article{Murk:2024nod,
    author = "Murk, Sebastian and Soranidis, Ioannis",
    title = "{Light rings and causality for nonsingular ultracompact objects sourced by nonlinear electrodynamics}",
    eprint = "2406.07957",
    archivePrefix = "arXiv",
    primaryClass = "gr-qc",
    doi = "10.1103/PhysRevD.110.044064",
    journal = "Phys. Rev. D",
    volume = "110",
    number = "4",
    pages = "044064",
    year = "2024"
}

@article{Bronnikov:2000vy,
    author = "Bronnikov, Kirill A.",
    title = "{Regular magnetic black holes and monopoles from nonlinear electrodynamics}",
    eprint = "gr-qc/0006014",
    archivePrefix = "arXiv",
    doi = "10.1103/PhysRevD.63.044005",
    journal = "Phys. Rev. D",
    volume = "63",
    pages = "044005",
    year = "2001"
}

@article{Pellicer:1969cf,
    author = "Pellicer, R. and Torrence, R. J.",
    title = "{Nonlinear electrodynamics and general relativity}",
    doi = "10.1063/1.1665019",
    journal = "J. Math. Phys.",
    volume = "10",
    pages = "1718--1723",
    year = "1969"
}

@article{Bakopoulos:2023fmv,
    author = "Bakopoulos, Athanasios and Charmousis, Christos and Kanti, Panagiota and Lecoeur, Nicolas and Nakas, Theodoros",
    title = "{Black holes with primary scalar hair}",
    eprint = "2310.11919",
    archivePrefix = "arXiv",
    primaryClass = "gr-qc",
    doi = "10.1103/PhysRevD.109.024032",
    journal = "Phys. Rev. D",
    volume = "109",
    number = "2",
    pages = "024032",
    year = "2024"
}

@article{Bakopoulos:2023sdm,
    author = "Bakopoulos, Athanasios and Chatzifotis, Nikos and Nakas, Theodoros",
    title = "{Compact objects with primary hair in shift and parity symmetric beyond Horndeski gravities}",
    eprint = "2312.17198",
    archivePrefix = "arXiv",
    primaryClass = "gr-qc",
    doi = "10.1103/PhysRevD.110.024044",
    journal = "Phys. Rev. D",
    volume = "110",
    number = "2",
    pages = "024044",
    year = "2024"
}

@article{Martinez:2004nb,
    author = "Martinez, Cristian and Troncoso, Ricardo and Zanelli, Jorge",
    title = "{Exact black hole solution with a minimally coupled scalar field}",
    eprint = "hep-th/0406111",
    archivePrefix = "arXiv",
    reportNumber = "CECS-PHY-04-12",
    doi = "10.1103/PhysRevD.70.084035",
    journal = "Phys. Rev. D",
    volume = "70",
    pages = "084035",
    year = "2004"
}

@article{bardeen1968,
    author = "J.M. Bardeen",
    title = "{Non-singular General Relativistic Gravitational Collapse}",
   journal = "Proceeding of the International Conference GR5, 174, Tbilisi University 
Press (1968)"
}

@article{Bokulic:2025brf,
    author = "Bokuli{\'c}, Ana and Juri{\'c}, Tajron and Smoli{\'c}, Ivica",
    title = "{Conundrum of regular black holes with nonlinear electromagnetic fields}",
    eprint = "2510.23711",
    archivePrefix = "arXiv",
    primaryClass = "gr-qc",
    reportNumber = "ZTF-EP-25-07; RBI-ThPhys-2025-41",
    month = "10",
    year = "2025"
}

@article{Tsukamoto:2020bjm,
    author = "Tsukamoto, Naoki",
    title = "{Gravitational lensing in the Simpson-Visser black-bounce spacetime in a strong deflection limit}",
    eprint = "2011.03932",
    archivePrefix = "arXiv",
    primaryClass = "gr-qc",
    doi = "10.1103/PhysRevD.103.024033",
    journal = "Phys. Rev. D",
    volume = "103",
    number = "2",
    pages = "024033",
    year = "2021"
}

@article{Morris:1988cz,
    author = "Morris, M. S. and Thorne, K. S.",
    title = "{Wormholes in space-time and their use for interstellar travel: A tool for teaching general relativity}",
    doi = "10.1119/1.15620",
    journal = "Am. J. Phys.",
    volume = "56",
    pages = "395--412",
    year = "1988"
}

@article{Correa:2013bza,
    author = "Correa, Francisco and Hassaine, Mokhtar",
    title = "{Thermodynamics of Lovelock black holes with a nonminimal scalar field}",
    eprint = "1312.4516",
    archivePrefix = "arXiv",
    primaryClass = "hep-th",
    doi = "10.1007/JHEP02(2014)014",
    journal = "JHEP",
    volume = "02",
    pages = "014",
    year = "2014"
}

@article{Altamirano:2014tva,
    author = "Altamirano, Natacha and Kubiznak, David and Mann, Robert B. and Sherkatghanad, Zeinab",
    title = "{Thermodynamics of rotating black holes and black rings: phase transitions and thermodynamic volume}",
    eprint = "1401.2586",
    archivePrefix = "arXiv",
    primaryClass = "hep-th",
    doi = "10.3390/galaxies2010089",
    journal = "Galaxies",
    volume = "2",
    pages = "89--159",
    year = "2014"
}

@article{Ahmed:2026bwm,
    author = "Ahmed, Faizuddin and Al-Badawi, Ahmad and Fathi, Mohsen",
    title = "{Charged Simpson-Visser AdS Black Holes: Geodesic Structure and Thermodynamic Properties}",
    eprint = "2601.10469",
    archivePrefix = "arXiv",
    primaryClass = "gr-qc",
    month = "1",
    year = "2026"
}

@article{Yu:2026syo,
    author = "Yu, Peng and Zhong, Yuan",
    title = "{First Law for Nonsingular Black Holes in 2D Dilaton Gravity}",
    eprint = "2603.21186",
    archivePrefix = "arXiv",
    primaryClass = "gr-qc",
    month = "3",
    year = "2026"
}

@article{Borde:1996df,
    author = "Borde, Arvind",
    title = "{Regular black holes and topology change}",
    eprint = "gr-qc/9612057",
    archivePrefix = "arXiv",
    primaryClass = "gr-qc",
    doi = "10.1103/PhysRevD.55.7615",
    journal = "Phys. Rev. D",
    volume = "55",
    pages = "7615-7617",
    year = "1997"
}

@article{Burinskii:2002pz,
    author = "Burinskii, A. and Hildebrandt, S. R.",
    title = "{New type of regular black holes and particle - like solutions from 
NED}",
    eprint = "hep-th/0202066",
    archivePrefix = "arXiv",
    primaryClass = "hep-th",
    doi = "10.1103/PhysRevD.65.104017",
    journal = "Phys. Rev. D",
    volume = "65",
    pages = "104017",
    year = "2002"
}

@article{Hayward:2005gi,
    author = "Hayward, Sean A.",
    title = "{Formation and evaporation of regular black holes}",
    eprint = "gr-qc/0506126",
    archivePrefix = "arXiv",
    primaryClass = "gr-qc",
    doi = "10.1103/PhysRevLett.96.031103",
    journal = "Phys. Rev. Lett.",
    volume = "96",
    pages = "031103",
    year = "2006"
}

@article{Berej:2006cc,
    author = "Berej, W. and Matyjasek, J. and Tryniecki, D. and Woronowicz, M.",
    title = "{Regular black holes in quadratic gravity}",
    eprint = "hep-th/0606185",
    archivePrefix = "arXiv",
    primaryClass = "hep-th",
    doi = "10.1007/s10714-006-0270-9",
    journal = "Gen. Rel. Grav.",
    volume = "38",
    pages = "885-906",
    year = "2006"
}

@article{Bronnikov:2006fu,
    author = "Bronnikov, K. A. and Melnikov, V. N. and Dehnen, H.",
    title = "{Regular black holes and black universes}",
    eprint = "gr-qc/0611022",
    archivePrefix = "arXiv",
    primaryClass = "gr-qc",
    doi = "10.1007/s10714-007-0430-6",
    journal = "Gen. Rel. Grav.",
    volume = "39",
    pages = "973-987",
    year = "2007"
}

@article{Lemos:2011dq,
    author = "Lemos, Jose P. S. and Zanchin, Vilson T.",
    title = "{Regular black holes: Electrically charged solutions, 
Reissner-Nordstr{\"o}m outside a de Sitter core}",
    eprint = "1104.4790",
    archivePrefix = "arXiv",
    primaryClass = "gr-qc",
    doi = "10.1103/PhysRevD.83.124005",
    journal = "Phys. Rev. D",
    volume = "83",
    pages = "124005",
    year = "2011"
}

@article{Flachi:2012nv,
    author = "Flachi, Antonino and Lemos, Jose P. S.",
    title = "{Quasinormal modes of regular black holes}",
    eprint = "1211.6212",
    archivePrefix = "arXiv",
    primaryClass = "gr-qc",
    doi = "10.1103/PhysRevD.87.024034",
    journal = "Phys. Rev. D",
    volume = "87",
    number = "2",
    pages = "024034",
    year = "2013"
}

@article{Li:2013jra,
    author = "Li, Zilong and Bambi, Cosimo",
    title = "{Measuring the Kerr spin parameter of regular black holes from 
their shadow}",
    eprint = "1309.1606",
    archivePrefix = "arXiv",
    primaryClass = "gr-qc",
    doi = "10.1088/1475-7516/2014/01/041",
    journal = "JCAP",
    volume = "01",
    pages = "041",
    year = "2014"
}

@article{Bambi:2013ufa,
    author = "Bambi, Cosimo and Modesto, Leonardo",
    title = "{Rotating regular black holes}",
    eprint = "1302.6075",
    archivePrefix = "arXiv",
    primaryClass = "gr-qc",
    doi = "10.1016/j.physletb.2013.03.025",
    journal = "Phys. Lett. B",
    volume = "721",
    pages = "329-334",
    year = "2013"
}

@article{Neves:2014aba,
    author = "Neves, J. C. S. and Saa, Alberto",
    title = "{Regular rotating black holes and the weak energy condition}",
    eprint = "1402.2694",
    archivePrefix = "arXiv",
    primaryClass = "gr-qc",
    doi = "10.1016/j.physletb.2014.05.026",
    journal = "Phys. Lett. B",
    volume = "734",
    pages = "44-48",
    year = "2014"
}

@article{Abdujabbarov:2016hnw,
    author = "Abdujabbarov, Ahmadjon and Amir, Muhammed and Ahmedov, Bobomurat 
and Ghosh, Sushant G.",
    title = "{Shadow of rotating regular black holes}",
    eprint = "1604.03809",
    archivePrefix = "arXiv",
    primaryClass = "gr-qc",
    doi = "10.1103/PhysRevD.93.104004",
    journal = "Phys. Rev. D",
    volume = "93",
    number = "10",
    pages = "104004",
    year = "2016"
}

@article{Toshmatov:2017zpr,
    author = "Toshmatov, Bobir and Stuchl{\'\i}k, Zdeněk and Ahmedov, 
Bobomurat",
    title = "{Generic rotating regular black holes in general relativity coupled 
to nonlinear electrodynamics}",
    eprint = "1704.07300",
    archivePrefix = "arXiv",
    primaryClass = "gr-qc",
    doi = "10.1103/PhysRevD.95.084037",
    journal = "Phys. Rev. D",
    volume = "95",
    number = "8",
    pages = "084037",
    year = "2017"
}

@article{Jusufi:2018jof,
    author = "Jusufi, Kimet and {\"O}vg{\"u}n, Ali and Saavedra, Joel and 
V{\'a}squez, Yerko and Gonz{\'a}lez, P. A.",
    title = "{Deflection of light by rotating regular black holes using the 
Gauss-Bonnet theorem}",
    eprint = "1804.00643",
    archivePrefix = "arXiv",
    primaryClass = "gr-qc",
    doi = "10.1103/PhysRevD.97.124024",
    journal = "Phys. Rev. D",
    volume = "97",
    number = "12",
    pages = "124024",
    year = "2018"
}

@article{Carballo-Rubio:2018pmi,
    author = "Carballo-Rubio, Ra{\'u}l and Di Filippo, Francesco and Liberati, 
Stefano and Pacilio, Costantino and Visser, Matt",
    title = "{On the viability of regular black holes}",
    eprint = "1805.02675",
    archivePrefix = "arXiv",
    primaryClass = "gr-qc",
    doi = "10.1007/JHEP07(2018)023",
    journal = "JHEP",
    volume = "07",
    pages = "023",
    year = "2018"
}

@article{Ovgun:2019wej,
    author = "{\"O}vg{\"u}n, Ali",
    title = "{Weak field deflection angle by regular black holes with cosmic 
strings using the Gauss-Bonnet theorem}",
    eprint = "1902.04411",
    archivePrefix = "arXiv",
    primaryClass = "gr-qc",
    doi = "10.1103/PhysRevD.99.104075",
    journal = "Phys. Rev. D",
    volume = "99",
    number = "10",
    pages = "104075",
    year = "2019"
}

@article{Simpson:2019mud,
    author = "Simpson, Alex and Visser, Matt",
    title = "{Regular black holes with asymptotically Minkowski cores}",
    eprint = "1911.01020",
    archivePrefix = "arXiv",
    primaryClass = "gr-qc",
    doi = "10.3390/universe6010008",
    journal = "Universe",
    volume = "6",
    number = "1",
    pages = "8",
    year = "2019"
}

@article{Kumar:2019pjp,
    author = "Kumar, Rahul and Ghosh, Sushant G. and Wang, Anzhong",
    title = "{Shadow cast and deflection of light by charged rotating regular 
black holes}",
    eprint = "1912.05154",
    archivePrefix = "arXiv",
    primaryClass = "gr-qc",
    doi = "10.1103/PhysRevD.100.124024",
    journal = "Phys. Rev. D",
    volume = "100",
    number = "12",
    pages = "124024",
    year = "2019"
}

@article{Jusufi:2022rbt,
    author = "Jusufi, Kimet",
    title = "{Regular solutions for black strings and torus-like black holes}",
    eprint = "2212.06760",
    archivePrefix = "arXiv",
    primaryClass = "gr-qc",
    doi = "10.1016/j.dark.2022.101156",
    journal = "Phys. Dark Univ.",
    volume = "39",
    pages = "101156",
    year = "2023"
}

@article{Bueno:2024dgm,
    author = "Bueno, Pablo and Cano, Pablo A. and Hennigar, Robie A.",
    title = "{Regular black holes from pure gravity}",
    eprint = "2403.04827",
    archivePrefix = "arXiv",
    primaryClass = "gr-qc",
    doi = "10.1016/j.physletb.2025.139260",
    journal = "Phys. Lett. B",
    volume = "861",
    pages = "139260",
    year = "2025"
}

@article{Dymnikova:2004zc,
    author = "Dymnikova, Irina",
    title = "{Regular electrically charged structures in nonlinear 
electrodynamics coupled to general relativity}",
    eprint = "gr-qc/0407072",
    archivePrefix = "arXiv",
    primaryClass = "gr-qc",
    doi = "10.1088/0264-9381/21/18/009",
    journal = "Class. Quant. Grav.",
    volume = "21",
    pages = "4417-4429",
    year = "2004"
}

@article{Nashed:2025bxv,
    author = "Nashed, G. G. L. and Saridakis, Emmanuel N.",
    title = "{3-dimensional charged black holes in $f({Q})$ gravity}",
    eprint = "2506.10046",
    archivePrefix = "arXiv",
    primaryClass = "gr-qc",
    month = "6",
    year = "2025"
}

@article{Nashed:2022yfc,
    author = "Nashed, G. G. L. and Saridakis, Emmanuel N.",
    title = "{New anisotropic star solutions in mimetic gravity}",
    eprint = "2206.12256",
    archivePrefix = "arXiv",
    primaryClass = "gr-qc",
    doi = "10.1140/epjp/s13360-023-03767-y",
    journal = "Eur. Phys. J. Plus",
    volume = "138",
    pages = "318",
    year = "2023"
}

@article{Nashed:2021pah,
    author = "Nashed, G. G. L. and Saridakis, Emmanuel N.",
    title = "{Stability of motion and thermodynamics in charged black holes in 
f(T) gravity}",
    eprint = "2111.06359",
    archivePrefix = "arXiv",
    primaryClass = "gr-qc",
    doi = "10.1088/1475-7516/2022/05/017",
    journal = "JCAP",
    volume = "05",
    number = "05",
    pages = "017",
    year = "2022"
}

@article{Nashed:2020kdb,
    author = "Nashed, G. G. L. and Saridakis, Emmanuel N.",
    title = "{New rotating black holes in nonlinear Maxwell $f({\mathcal R})$ 
gravity}",
    eprint = "2010.10422",
    archivePrefix = "arXiv",
    primaryClass = "gr-qc",
    reportNumber = "Phys.Rev.D 102 (2020) 124072",
    doi = "10.1103/PhysRevD.102.124072",
    journal = "Phys. Rev. D",
    volume = "102",
    number = "12",
    pages = "124072",
    year = "2020"
}

@article{Nashed:2018cth,
    author = "Nashed, G. G. L. and Saridakis, Emmanuel N.",
    title = "{Rotating AdS black holes in Maxwell-$f(T)$ gravity}",
    eprint = "1811.03658",
    archivePrefix = "arXiv",
    primaryClass = "gr-qc",
    doi = "10.1088/1361-6382/ab23d9",
    journal = "Class. Quant. Grav.",
    volume = "36",
    number = "13",
    pages = "135005",
    year = "2019"
}

@article{Cognola:2011nj,
    author = "Cognola, Guido and Gorbunova, Olesya and Sebastiani, Lorenzo and 
Zerbini, Sergio",
    title = "{On the Energy Issue for a Class of Modified Higher Order Gravity 
Black Hole Solutions}",
    eprint = "1104.2814",
    archivePrefix = "arXiv",
    primaryClass = "gr-qc",
    doi = "10.1103/PhysRevD.84.023515",
    journal = "Phys. Rev. D",
    volume = "84",
    pages = "023515",
    year = "2011"
}

@article{Myung:2013doa,
    author = "Myung, Yun Soo",
    title = "{Stability of Schwarzschild black holes in fourth-order gravity 
revisited}",
    eprint = "1306.3725",
    archivePrefix = "arXiv",
    primaryClass = "gr-qc",
    doi = "10.1103/PhysRevD.88.024039",
    journal = "Phys. Rev. D",
    volume = "88",
    number = "2",
    pages = "024039",
    year = "2013"
}

@article{Bambi:2015kza,
    author = "Bambi, Cosimo",
    title = "{Testing black hole candidates with electromagnetic radiation}",
    eprint = "1509.03884",
    archivePrefix = "arXiv",
    primaryClass = "gr-qc",
    doi = "10.1103/RevModPhys.89.025001",
    journal = "Rev. Mod. Phys.",
    volume = "89",
    number = "2",
    pages = "025001",
    year = "2017"
}

@article{Nozari:2025rkc,
    author = "Nozari, Kourosh and Saghafi, Sara and Ramezanpasandi, Zeynab",
    title = "{Accretion process as a probe of extra dimensions in MOG compact 
object spacetimes}",
    eprint = "2511.01677",
    archivePrefix = "arXiv",
    primaryClass = "gr-qc",
    doi = "10.1140/epjc/s10052-025-14915-2",
    journal = "Eur. Phys. J. C",
    volume = "85",
    number = "10",
    pages = "1173",
    year = "2025"
}

@article{Bueno:2025gjg,
    author = "Bueno, Pablo and Cano, Pablo A. and Hennigar, Robie A. and Murcia, 
{\'A}ngel J. and Vicente-Cano, Aitor",
    title = "{Regular black holes from Oppenheimer-Snyder collapse}",
    eprint = "2505.09680",
    archivePrefix = "arXiv",
    primaryClass = "gr-qc",
    doi = "10.1103/qrbb-mdvm",
    journal = "Phys. Rev. D",
    volume = "112",
    number = "6",
    pages = "064039",
    year = "2025"
}

@article{Charmousis:2025xug,
    author = "Charmousis, Christos and Iteanu, Simon and Langlois, David and 
Noui, Karim",
    title = "{Axial perturbations of black holes with primary scalar hair}",
    eprint = "2503.22348",
    archivePrefix = "arXiv",
    primaryClass = "gr-qc",
    reportNumber = "CERN-TH-2025-058",
    doi = "10.1088/1475-7516/2025/05/102",
    journal = "JCAP",
    volume = "05",
    pages = "102",
    year = "2025"
}

@article{Nozari:2024vxp,
    author = "Nozari, Kourosh and Saghafi, Sara and Hassani, Mohammad",
    title = "{Accretion onto a charged black hole in consistent 4D 
Einstein-Gauss-Bonnet gravity}",
    eprint = "2412.07814",
    archivePrefix = "arXiv",
    primaryClass = "gr-qc",
    doi = "10.1016/j.jheap.2024.12.004",
    journal = "JHEAp",
    volume = "45",
    pages = "214--230",
    year = "2025"
}

@article{Barrientos:2022bzm,
    author = "Barrientos, Jose and Cisterna, Adolfo and Kubiznak, David and 
Oliva, Julio",
    title = "{Accelerated black holes beyond Maxwell's electrodynamics}",
    eprint = "2205.15777",
    archivePrefix = "arXiv",
    primaryClass = "gr-qc",
    doi = "10.1016/j.physletb.2022.137447",
    journal = "Phys. Lett. B",
    volume = "834",
    pages = "137447",
    year = "2022"
}

@article{Barrientos:2024umq,
    author = "Barrientos, Jos{\'e} and Cisterna, Adolfo and Hassaine, Mokhtar 
and Pallikaris, Konstantinos",
    title = "{Electromagnetized black holes and swirling backgrounds in 
nonlinear electrodynamics: The ModMax case}",
    eprint = "2409.12336",
    archivePrefix = "arXiv",
    primaryClass = "gr-qc",
    doi = "10.1016/j.physletb.2024.139214",
    journal = "Phys. Lett. B",
    volume = "860",
    pages = "139214",
    year = "2025"
}

@article{Cisterna:2020rkc,
    author = "Cisterna, Adolfo and Giribet, Gaston and Oliva, Julio and 
Pallikaris, Konstantinos",
    title = "{Quasitopological electromagnetism and black holes}",
    eprint = "2004.05474",
    archivePrefix = "arXiv",
    primaryClass = "hep-th",
    doi = "10.1103/PhysRevD.101.124041",
    journal = "Phys. Rev. D",
    volume = "101",
    number = "12",
    pages = "124041",
    year = "2020"
}

@article{Barrientos:2025rjn,
    author = {Barrientos, Jos{\'e} and Cisterna, Adolfo and Hassaine, Mokhtar 
and M{\"u}ller, Keanu and Pallikaris, Konstantinos},
    title = "{A new exact rotating spacetime in vacuum: The 
Kerr{\textendash}Levi-Civita spacetime}",
    eprint = "2506.07166",
    archivePrefix = "arXiv",
    primaryClass = "gr-qc",
    doi = "10.1016/j.physletb.2025.140035",
    journal = "Phys. Lett. B",
    volume = "871",
    pages = "140035",
    year = "2025"
}

@article{Cisterna:2025vxk,
    author = "Cisterna, Adolfo and Hassaine, Mokhtar and Hernandez-Vera, 
Ulises",
    title = "{Thermodynamics of four-dimensional regular black holes with an 
infinite tower of regularized curvature corrections}",
    eprint = "2505.23467",
    archivePrefix = "arXiv",
    primaryClass = "gr-qc",
    doi = "10.1103/6f3b-8794",
    journal = "Phys. Rev. D",
    volume = "112",
    number = "6",
    pages = "064036",
    year = "2025"
}

@article{Babichev:2020qpr,
    author = "Babichev, Eugeny and Charmousis, Christos and Cisterna, Adolfo 
and 
Hassaine, Mokhtar",
    title = "{Regular black holes via the Kerr-Schild construction in DHOST 
theories}",
    eprint = "2004.00597",
    archivePrefix = "arXiv",
    primaryClass = "hep-th",
    doi = "10.1088/1475-7516/2020/06/049",
    journal = "JCAP",
    volume = "06",
    pages = "049",
    year = "2020"
}

@article{Boos:2024sgm,
    author = "Boos, Jens",
    title = "{What happens to topological invariants and black holes in 
singularity-free theories?}",
    eprint = "2411.11450",
    archivePrefix = "arXiv",
    primaryClass = "gr-qc",
    doi = "10.1103/PhysRevD.111.084063",
    journal = "Phys. Rev. D",
    volume = "111",
    number = "8",
    pages = "084063",
    year = "2025"
}

@article{Estrada:2024uuu,
    author = "Estrada, Milko and Aros, Rodrigo",
    title = "{Pure Lovelock gravity regular black holes}",
    eprint = "2409.09559",
    archivePrefix = "arXiv",
    primaryClass = "gr-qc",
    doi = "10.1088/1475-7516/2025/01/032",
    journal = "JCAP",
    volume = "01",
    pages = "032",
    year = "2025"
}

@article{Junior:2023ixh,
    author = "Junior, Jos{\'e} Tarciso S. S. and Lobo, Francisco S. N. and 
Rodrigues, Manuel E.",
    title = "{(Regular) Black holes in conformal Killing gravity coupled to 
nonlinear electrodynamics and scalar fields}",
    eprint = "2310.19508",
    archivePrefix = "arXiv",
    primaryClass = "gr-qc",
    doi = "10.1088/1361-6382/ad210e",
    journal = "Class. Quant. Grav.",
    volume = "41",
    number = "5",
    pages = "055012",
    year = "2024"
}

@article{Babichev:2023mgk,
    author = "Babichev, Eugeny and Charmousis, Christos and Lecoeur, Nicolas",
    title = "{Rotating black holes embedded in a cosmological background for 
scalar-tensor theories}",
    eprint = "2305.17129",
    archivePrefix = "arXiv",
    primaryClass = "gr-qc",
    doi = "10.1088/1475-7516/2023/08/022",
    journal = "JCAP",
    volume = "08",
    pages = "022",
    year = "2023"
}

@article{Nozari:2023enj,
    author = "Nozari, Kourosh and Saghafi, Sara and Aliyan, Fateme",
    title = "{Accretion onto a static spherically symmetric regular MOG dark 
compact object}",
    eprint = "2305.17186",
    archivePrefix = "arXiv",
    primaryClass = "gr-qc",
    doi = "10.1140/epjc/s10052-023-11620-w",
    journal = "Eur. Phys. J. C",
    volume = "83",
    number = "5",
    pages = "449",
    year = "2023"
}

@article{Konoplya:2023ppx,
    author = "Konoplya, R. A.",
    title = "{Quasinormal modes and grey-body factors of regular black holes 
with a scalar hair from the Effective Field Theory}",
    eprint = "2305.09187",
    archivePrefix = "arXiv",
    primaryClass = "gr-qc",
    doi = "10.1088/1475-7516/2023/07/001",
    journal = "JCAP",
    volume = "07",
    pages = "001",
    year = "2023"
}

@article{Carballo-Rubio:2021wjq,
    author = "Carballo-Rubio, Ra{\'u}l and Di Filippo, Francesco and Liberati, 
Stefano and Visser, Matt",
    title = "{Geodesically complete black holes in Lorentz-violating gravity}",
    eprint = "2111.03113",
    archivePrefix = "arXiv",
    primaryClass = "gr-qc",
    doi = "10.1007/JHEP02(2022)122",
    journal = "JHEP",
    volume = "02",
    pages = "122",
    year = "2022"
}

@article{Sucu:2026nkw,
    author = "Sucu, Erdem and Sakalli, Izzet and Saridakis, Emmanuel N.",
    title = "{Branch structure and nonextensive thermodynamics of 
Kalb-Ramond-ModMax black holes: observational signatures}",
    eprint = "2601.20004",
    archivePrefix = "arXiv",
    primaryClass = "gr-qc",
    month = "1",
    year = "2026"
}

@article{Anand:2025cer,
    author = "Anand, Ankit and Jusufi, Kimet and Basilakos, Spyros and 
Saridakis, Emmanuel N.",
    title = "{Topological and optical signatures of modified black-hole 
entropies}",
    eprint = "2512.13769",
    archivePrefix = "arXiv",
    primaryClass = "gr-qc",
    doi = "10.1140/epjc/s10052-026-15365-0",
    journal = "Eur. Phys. J. C",
    volume = "86",
    number = "2",
    pages = "126",
    year = "2026"
}

@article{Nozari:2026wjo,
    author = "Nozari, Kourosh and Hajebrahimi, Milad and Saghafi, Sara and 
Mustafa, G. and Saridakis, Emmanuel N.",
    title = "{Rotating Black Holes with Primary Scalar Hair: Shadow Signatures 
in Beyond Horndeski Gravity}",
    eprint = "2602.16237",
    archivePrefix = "arXiv",
    primaryClass = "gr-qc",
    month = "2",
    year = "2026"
}

@article{Santos:2025fdp,
    author = "Santos, Fabiano F. and Pourhassan, Behnam and Saridakis, Emmanuel 
N.",
    title = "{Black Hole Entropy and Complexity Growth in Horndeski Gravity 
within the AdS/BCFT Framework}",
    eprint = "2509.23430",
    archivePrefix = "arXiv",
    primaryClass = "hep-th",
    month = "9",
    year = "2025"
}

@article{Ali:2018boy,
    author = "Ali, Md. Sabir and Ghosh, Sushant G.",
    title = "{Exact $d$-dimensional Bardeen-de Sitter black holes and 
thermodynamics}",
    doi = "10.1103/PhysRevD.98.084025",
    journal = "Phys. Rev. D",
    volume = "98",
    number = "8",
    pages = "084025",
    year = "2018"
}

@article{Luciano:2023bai,
    author = "Luciano, Giuseppe Gaetano and Saridakis, Emmanuel",
    title = "{P {\ensuremath{-}} v criticalities, phase transitions and 
geometrothermodynamics of charged AdS black holes from Kaniadakis statistics}",
    eprint = "2308.12669",
    archivePrefix = "arXiv",
    primaryClass = "gr-qc",
    doi = "10.1007/JHEP12(2023)114",
    journal = "JHEP",
    volume = "12",
    pages = "114",
    year = "2023"
}

@article{Johannsen:2013szh,
    author = "Johannsen, Tim",
    title = "{Regular Black Hole Metric with Three Constants of Motion}",
    eprint = "1501.02809",
    archivePrefix = "arXiv",
    primaryClass = "gr-qc",
    doi = "10.1103/PhysRevD.88.044002",
    journal = "Phys. Rev. D",
    volume = "88",
    number = "4",
    pages = "044002",
    year = "2013"
}

@article{Vagnozzi:2022moj,
    author = "Vagnozzi, Sunny and others",
    title = "{Horizon-scale tests of gravity theories and fundamental physics from the Event Horizon Telescope image of Sagittarius A}",
    eprint = "2205.07787",
    archivePrefix = "arXiv",
    primaryClass = "gr-qc",
    reportNumber = "UCI-HEP-TR-2022-07",
    doi = "10.1088/1361-6382/acd97b",
    journal = "Class. Quant. Grav.",
    volume = "40",
    number = "16",
    pages = "165007",
    year = "2023"
}

@article{Calza:2024fzo,
    author = "Calz{\`a}, Marco and Pedrotti, Davide and Vagnozzi, Sunny",
    title = "{Primordial regular black holes as all the dark matter. I. Time-radial-symmetric metrics}",
    eprint = "2409.02804",
    archivePrefix = "arXiv",
    primaryClass = "gr-qc",
    doi = "10.1103/PhysRevD.111.024009",
    journal = "Phys. Rev. D",
    volume = "111",
    number = "2",
    pages = "024009",
    year = "2025"
}

@article{Calza:2025mwn,
    author = "Calz{\`a}, Marco and Pedrotti, Davide and Yuan, Guan-Wen and Vagnozzi, Sunny",
    title = "{Primordial regular black holes as all the dark matter. III. Covariant canonical quantum gravity models}",
    eprint = "2507.02396",
    archivePrefix = "arXiv",
    primaryClass = "gr-qc",
    doi = "10.1103/4x1f-vctx",
    journal = "Phys. Rev. D",
    volume = "112",
    number = "12",
    pages = "124015",
    year = "2025"
}

@article{Calza:2024xdh,
    author = "Calz{\`a}, Marco and Pedrotti, Davide and Vagnozzi, Sunny",
    title = "{Primordial regular black holes as all the dark matter. II. Non-time-radial-symmetric and loop quantum gravity-inspired metrics}",
    eprint = "2409.02807",
    archivePrefix = "arXiv",
    primaryClass = "gr-qc",
    doi = "10.1103/PhysRevD.111.024010",
    journal = "Phys. Rev. D",
    volume = "111",
    number = "2",
    pages = "024010",
    year = "2025"
}

@article{Calza:2024qxn,
    author = "Calz{\`a}, Marco and Gianesello, Francesco and Rinaldi, Massimiliano and Vagnozzi, Sunny",
    title = "{Implications of cosmologically coupled black holes for pulsar timing arrays}",
    eprint = "2409.01801",
    archivePrefix = "arXiv",
    primaryClass = "gr-qc",
    doi = "10.1038/s41598-024-82661-8",
    journal = "Sci. Rep.",
    volume = "14",
    number = "1",
    pages = "31296",
    year = "2024"
}

@article{Calza:2025yfm,
    author = "Calz{\`a}, Marco and Rinaldi, Massimiliano and Vagnozzi, Sunny",
    title = "{Importance of being nonminimally coupled: Scalar Hawking radiation from regular black holes}",
    eprint = "2510.12257",
    archivePrefix = "arXiv",
    primaryClass = "gr-qc",
    doi = "10.1103/s4cf-bclq",
    journal = "Phys. Rev. D",
    volume = "112",
    number = "10",
    pages = "104055",
    year = "2025"
}
\bibliographystyle{utphys}

\end{document}